\documentclass[11pt]{article}%
\usepackage{amsfonts,amssymb}
\usepackage[osf,sc]{mathpazo}
\usepackage{amsmath}
\usepackage{amsfonts}
\usepackage{amssymb}
\usepackage{graphicx}
\usepackage{hyperref}%
\providecommand{\U}[1]{\protect\rule{.1in}{.1in}}
\def\be{\begin{eqnarray}}
\def\ee{\end{eqnarray}}
\def\tanc {\rm tanc}
\def\openone {1\kern -0.36em\llap~1}
\begin{document}

\begin{center}
{\LARGE Umbral Vade Mecum}\medskip

{\Large Thomas L Curtright}$^{\S }$ and {\Large Cosmas K Zachos}$^{\natural}%
$\bigskip

$^{\S }$Department of Physics, University of Miami, Coral Gables, FL
33124-8046, USA\newline\textsl{curtright@miami.edu}

$^{\natural}$High Energy Physics Division, Argonne National Laboratory,
Argonne, IL 60439-4815, USA \newline\textsl{zachos@anl.gov}\bigskip
\end{center}

\centerline{\bf Abstract} \phantom{bunga} \textit{In recent years the umbral
calculus has emerged from the shadows\ to provide an elegant correspondence
framework that automatically gives systematic solutions of ubiquitous
difference equations --- discretized versions of the differential cornerstones
appearing in most areas of physics and engineering --- as maps of well-known
continuous functions. \ This correspondence deftly sidesteps the use of more
traditional methods to solve these difference equations. \ The umbral
framework is discussed and illustrated here, with special attention given to
umbral counterparts of the Airy, Kummer, and Whittaker equations, and to
umbral maps of solitons for the Sine-Gordon, Korteweg--de Vries, and Toda
systems.}
%\pacs{03.65.Fd,11.30.-j,04.60.Nc,04.60.Pp}

\section{Introduction}

Robust theoretical arguments have established an anticipation of a fundamental
minimum measurable length in Nature, of order $L_{_{\text{Planck}}}\equiv
\sqrt{\hbar G_{N}/c^{3}}=1.6162\times10^{-35}%
%TCIMACRO{\unit{m}}%
%BeginExpansion
\operatorname{m}%
%EndExpansion
$, the corresponding mass and time being $M_{_{\text{Planck}}}\equiv
\sqrt{\hbar c/G_{N}}=2.1765\times10^{-8}%
%TCIMACRO{\unit{kg}}%
%BeginExpansion
\operatorname{kg}%
%EndExpansion
$ and $L_{_{\text{Planck}}}/c=5.3911\times10^{-44}%
%TCIMACRO{\unit{s}}%
%BeginExpansion
\operatorname{s}%
%EndExpansion
$. \ The essence of such arguments is the following (in relativistic quantum
geometrical units, wherein $\hbar$, $c$, and $M_{_{\text{Planck}}}$ are all unity).

In a system or process characterized by energy $E$, no lengths smaller than
$L$ can be measured, where $L$ is the \emph{larger} of either the
Schwarzschild horizon radius of the system ($\sim E$) or, for energies smaller
than the Planck mass, the Compton wavelength of the aggregate process
($\sim1/E$). \ Since the minimum of $\max(E,1/E)$ lies at the Planck mass
($E=1$), the smallest measurable distance is widely believed to be of order
$L_{_{\text{Planck}}}$. \ Thus, continuum laws in Nature are expected to be
deformed, in principle, by modifications at that minimum length scale. \ 

Remarkably, however, if a fundamental spacetime lattice of spacing
$a=O(L_{_{PLanck}})$ is the structure that underlies conventional continuum
physics, then it turns out that continuous symmetries, such as Galilei or
Lorentz invariance, can actually survive unbroken under such a deformation
into discreteness, in a nonlocal, \emph{umbral realization} \cite{Smirnov,LTW}.

Umbral calculus, pioneered by Rota and associates in a combinatorial context
\cite{rota,loeb}, specifies, in principle, how functions of discrete variables
in \emph{infinite} domains provide systematic \textquotedblleft
shadows\textquotedblright\ of their familiar continuum limit properties. \ By
preserving Leibniz's chain rule, and by providing a discrete counterpart of
the Heisenberg algebra, observables built from difference operators shadow the
Lie algebras of the standard differential operators of continuum physics.
\ (For a review relevant to physics, see \cite{levi}.) \ Nevertheless, while
the continuous symmetries and Lie algebras of umbrally deformed systems might
remain identical to their continuum limit, the functions of observables
themselves are modified, in general, and often drastically so. \ 

\pagestyle{myheadings} \markright{\small{\sf Curtright \& Zachos --- Umbral Vade Mecum --- arXiv:1304.0429  \qquad  }}

Traditionally, the controlling continuum differential equations of physics are
first discretized \cite{benderorszag,Smirnov,dimakis}, and then those
difference equations are solved to yield umbral deformations of the continuum
solutions. \ But quite often, routine methods to solve such discrete equations
become unwieldy, if not intractable. \ On the other hand, some technical
difficulties may be bypassed by directly discretizing the continuum solutions.
\ That is, through appropriate umbral deformation of the continuum solutions,
the corresponding discrete difference equations may be automatically solved.
\ However, as illustrated below for the simplest cases of oscillations and
wave propagation, the resulting umbral modifications may present some
subtleties when it comes to extracting the underlying physics. \ 

In \cite{ckz} the linearity of the umbral deformation functional was
exploited, together with the fact that the umbral image of an exponential is
also an exponential,\ albeit with interesting modifications, to discretize
well-behaved functions occurring in \emph{solutions} of physical differential
equations through their Fourier expansion. \ This discrete shadowing of the
Fourier representation functional should thus be of utility in inferring wave
disturbance propagation in discrete spacetime lattices. \ We continue to
pursue this idea here with some explicit examples. \ We do this in conjunction
with the umbral deformation of power series, especially those for
hypergeometric functions. \ We compare both Fourier and power series methods
in some detail to gain further insight into the umbral framework.

Overall, we utilize essentially all aspects of the elegant umbral calculus to
provide systematic solutions of discretized cornerstone differential equations
that are ubiquitous in most areas of physics and engineering. \ We pay
particular attention to the umbral counterparts of the Airy, Kummer, and
Whittaker equations, and their solutions, and to the umbral maps of solitons
for the Sine-Gordon, Korteweg--de Vries, and Toda systems.

\section{Overview of the umbral correspondence}

For simplicity, consider discrete time, $t=0,\ a,\ 2a,\ \cdots,\ na,\ \cdots$.
\ Without loss of generality, broadly following the summary review of
\cite{levi}, consider an umbral deformation defined by the forward difference
discretization of $\partial_{t}$,
\begin{equation}
\Delta x(t)\equiv\frac{x(t+a)-x(t)}{a}~, \label{plaindef}%
\end{equation}
and whence of the elementary oscillation equation, $\ddot{x}(t)=-x(t)$,
namely,
\begin{equation}
\Delta^{2}x(t)=\frac{x(t+2a)-2x(t+a)+x(t)}{a^{2}}=-x(t)~. \label{diffeqn}%
\end{equation}
Now consider the solutions of this second-order difference equation. \ Of
course, (\ref{diffeqn}) can be easily solved directly by the textbook
Fourier-component Ansatz $x(t)\propto r^{t}$, \cite{benderorszag}, to yield
$(1\pm ia)^{t/a}$. \ However, to illustrate instead the powerful systematics
of umbral calculus \cite{Smirnov,levi}, we produce and study the solution in
that framework.

The umbral framework considers associative chains of operators, generalizing
ordinary continuum functions by ultimately acting on a
translationally-invariant \textquotedblleft vacuum\textquotedblright\ $1$,
after manipulations to move shift operators to the right and have them
absorbed by that vacuum, which we indicate by $T\cdot1=1$. \ Using the
standard Lagrange-Boole shift generator
\begin{equation}
T\equiv~e^{a\partial_{t}},\qquad\text{\hbox{so that}}\qquad Tf(t)\cdot
1=f(t+a)~T\cdot1=f(t+a)~1,
\end{equation}
the umbral deformation is then
\begin{equation}
\partial_{t}\qquad\longmapsto\qquad\Delta\equiv\frac{T-1}{a}~,
\end{equation}%
\begin{equation}
t\qquad\longmapsto\qquad tT^{-1},
\end{equation}%
\begin{equation}
t^{n}\qquad\longmapsto\qquad(tT^{-1})^{n}=t(t-a)(t-2a)...(t-(n-1)a)T^{-n}%
\equiv\lbrack t]^{n}T^{-n}, \label{[t]^n}%
\end{equation}
so that $[t]^{0}=1$, and, for $n>0$, $[0]^{n}=0$. \ The $[t]^{n}$ are called
\textquotedblleft basic polynomials\textquotedblright\footnote{We stress that
the notation $[t]^{n}$ is \emph{shorthand} for the product
$t(t-a)...(t-(n-1)a)$. \ It is \emph{not} just the $n$th power of $\left[
t\right]  =t$.} for positive $n$ \cite{rota,levi,dimakis}, and they are
eigenfunctions of $tT^{-1}\Delta$.

A linear combination of monomials (a power series representation of a
function) will thus transform umbrally to the same linear combination of basic
polynomials, with the same series coefficients, $f(t)~~\longmapsto
~~f(tT^{-1})$. \ All observables in the discretized world are thus such
deformation maps of the continuum observables, and evaluation of their direct
functional form is in order. \ Below, we will be concluding the correspondence
by casually eliminating translation operators at the very end, first through
operating on the vacuum and then leaving it implicit, so that $F\left(
t\right)  \equiv f(tT^{-1})\cdot1$.

The umbral deformation relies on the respective umbral entities obeying
operator combinatorics identical to their continuum limit ($a\rightarrow0$),
by virtue of obeying the \emph{same Heisenberg commutation relation}
\cite{Smirnov},
\begin{equation}
\lbrack\partial_{t},t]~=~1~=~[\Delta,tT^{-1}]~.
\end{equation}
Thus, e.g., by shift invariance, $T\Delta T^{-1}=\Delta$,
\begin{equation}
\lbrack\partial_{t},t^{n}]=nt^{n-1}\qquad\longmapsto\qquad\lbrack
{\Delta,[t]^{n}T^{-n}}]=n[t]^{n-1}~T^{1-n},
\end{equation}
so that, ultimately, $\Delta\lbrack t]^{n}=n[t]^{n-1}$. \ For commutators of
associative operators, the umbrally deformed Leibniz rule holds \cite{LTW},
\begin{equation}
\lbrack\Delta,f(tT^{-1})g(tT^{-1})]=[\Delta,f(tT^{-1})]g(tT^{-1}%
)+f(tT^{-1})[\Delta,g(tT^{-1})]~,
\end{equation}
ultimately to be dotted onto $1$
%Should we use [openone]<LaTeX>\openone</LaTeX> instead?
. \ Formally, the umbral deformation reflects (unitary) equivalences of the
unitary irreducible representation of the Heisenberg-Weyl group, provided for
by the Stone-von Neumann theorem. \ Here, these equivalences reflect the
alternate consistent realizations of all continuum physics structures through
systematic maps such as the one we have chosen. \ It is worth stressing that
the representations of this algebraic relation on the real or complex number
fields can only be \emph{infinite dimensional}, that is, the lattices covered
must be infinite. \ 

Now note that, in this case the basic polynomials $[t]^{n}$ are just scaled
falling factorials, for $n\geq0$, i.e. generalized Pochhammer symbols, which
may be expressed in various ways:
\begin{align}
\lbrack t]^{n}  &  \equiv\left(  tT^{-1}\right)  ^{n}\cdot1=t(t-a)\cdots
(t-\left(  n-1\right)  a)=a^{n}\frac{(t/a)!}{(t/a-n)!}\nonumber\\
&  =a^{n}\frac{\Gamma\left(  \frac{t}{a}+1\right)  }{\Gamma\left(  \frac{t}%
{a}-n+1\right)  }=\left(  -a\right)  ^{n}\frac{\Gamma\left(  n-\frac{t}%
{a}\right)  }{\Gamma\left(  -\frac{t}{a}\right)  }\ . \label{PosIntPower}%
\end{align}
Thus $[-t]^{n}=(-)^{n}[t+a(n-1)]^{n}$. \ Furthermore, $[an]^{n}=a^{n}n!$ ;
$[t]^{m}[t-am]^{n-m}=[t]^{n}$ for $0\leq m\leq n$ ; and for integers $0\leq
m<n$, $[am]^{n}=0$. \ Thus, $\Delta^{m}[t]^{n}=[an]^{m}[t]^{n-m}/a^{m}$.

Negative umbral powers, by contrast, are the inverse of rising factorials,
instead:
\begin{align}
\left[  \frac{1}{t}\right]  ^{n}  &  =\left(  T\frac{1}{t}\right)  ^{n}%
\cdot1={\frac{1}{(t+a)(t+2a)\cdots(t+na)}}=a^{-n}{\frac{(t/a)!}{(t/a+n)!}%
}\nonumber\\
&  =a^{-n}\frac{\Gamma\left(  \frac{t}{a}+1\right)  }{\Gamma\left(  \frac
{t}{a}+n+1\right)  }=\left(  -a\right)  ^{-n}\frac{\Gamma\left(  -\frac{t}%
{a}-n\right)  }{\Gamma\left(  -\frac{t}{a}\right)  }\ . \label{NegIntPower}%
\end{align}
These correspond to the negative eigenvalues of $tT^{-1}\Delta$.

The standard umbral exponential is then natural to define as
\cite{rota,floreanini}\footnote{Again we stress that $e^{\lambda\lbrack t]}$
is a short-hand notation, and not just the usual exponential of $\lambda
\left[  t\right]  =\lambda t$.}
\begin{equation}
E(\lambda t,\lambda a)\equiv e^{\lambda\lbrack t]}\equiv e^{\lambda tT^{-1}%
}\cdot1=\sum_{n=0}^{\infty}\frac{\lambda^{n}}{n!}[t]^{n}=\sum_{n=0}^{\infty
}(\lambda a)^{n}{\binom{t/a}{n}}=(1+\lambda a)^{t/a},
\end{equation}
the compound interest formula, with the proper continuum limit ($a\rightarrow
0$). \ N.B. \ There is always a 0 at $\lambda=-1/a$.

Evidently, since $\Delta\cdot1=0$,
\begin{equation}
\Delta e^{\lambda\lbrack t]}=\lambda~e^{\lambda\lbrack t]},
\end{equation}
and, as already indicated, one could have solved this equation
directly\footnote{N.B. \ There is an \emph{infinity} of
\emph{\textquotedblleft non-umbral\textquotedblright} extensions of the
$E\left(  \lambda t,\lambda a\right)  $ solution \cite{ltw}: \ Multiplying the
umbral exponential by an arbitrary periodic function $g(t+a)=g(t)$ will pass
undetected through $\Delta$, and thus will also yield an eigenfunction of
$\Delta$. \ Often, such extra solutions have either a vanishing continuum
limit, or else an ill-defined one.} to produce the above $E(\lambda t,\lambda
a)$.

Serviceably, the umbral exponential $E$ \emph{happens to be an ordinary
exponential},
\begin{equation}
e^{\lambda\lbrack t]}=e^{\frac{\ln(1+\lambda a)}{a}t}\ , \label{EisOrdinary}%
\end{equation}
and it actually serves as the generating function of the umbral basic
polynomials,
\begin{equation}
\frac{\partial^{n}}{\partial\lambda^{n}}(1+\lambda a)^{t/a}\Biggr |_{\lambda
=0}=[t]^{n}.
\end{equation}
Conversely, then, this construction may be reversed, by first solving directly
for the umbral eigenfunction of $\Delta$, and effectively defining the umbral
basic polynomials through the above parametric derivatives, in situations
where these might be more involved, as in the next section.

As a consequence of linearity, the umbral deformation of a power series
representation of a function is given formally by
\begin{equation}
f(t)~~\longmapsto~~F(t)\equiv f(tT^{-1})\cdot1=\left.  f\left(  \frac
{\partial}{\partial\lambda}\right)  ~(1+\lambda a)^{t/a}\right\vert
_{\lambda=0}\ . \label{foist}%
\end{equation}
This may not always be easy to evaluate, but, in fact, the same argument may
be applied to linear combinations of exponentials, and hence the entire
Fourier representation functional, to obtain
\begin{equation}
F(t)=\int_{-\infty}^{\infty}d\tau~f(\tau)\int_{-\infty}^{\infty}\frac{d\omega
}{2\pi}~e^{-i\omega\tau}(1+i\omega a)^{t/a}=\left.  \left(  1+a\frac{\partial
}{\partial\tau}\right)  ^{t/a}~f(\tau)\right\vert _{\tau=0}\ . \label{neato}%
\end{equation}
The rightmost equation follows by converting $i\omega$ into $\partial_{\tau}$
derivatives and integrating by parts away from the resulting delta function.
\ Naturally, it identifies with Eqn (\ref{foist}) by the (Fourier) identity
$f(\partial_{x})g(x)|_{x=0}=g(\partial_{x})f(x)|_{x=0}$. \ It is up to
individual ingenuity to utilize the form best suited to the particular
application at hand.

It is also straightforward to check that this umbral transform functional
yields
\begin{equation}
\partial_{t}f~\longmapsto~\Delta F\ ,
\end{equation}
and to evaluate the umbral transform of the Dirac delta function, which
amounts to a cardinal sine or sampling function,
\begin{equation}
\delta(t)~\longmapsto~{\frac{\sin({\frac{\pi}{2}}(1+t/a))}{(\pi(a+t))}}\ ,
\end{equation}
or to evaluate umbral transforms of rational functions, such as
\begin{equation}
f={\frac{1}{(1-t)}}~\longmapsto~F=e^{1/a}a^{t/a}\Gamma(t/a+1,1/a)\ ,
\end{equation}
to obtain an incomplete Gamma function
\href{http://www.convertit.com/Go/ConvertIt/Reference/AMS55.ASP?Res=150&Page=260}{(A\&S
6.5.3)}, and so on. \ Note how the last of these is distinctly, if subtly,
different from the umbral transform of negative powers, as given in
(\ref{NegIntPower}).

In practical applications, evaluation of umbral transforms of arbitrary
functions of observables may be more direct, at the level of solutions,
through this deforming functional, Eqn (\ref{neato}). \ For example, one may
evaluate in this way the umbral correspondents of trigonometric functions,
\begin{equation}
\operatorname{Sin}[t]\equiv\frac{e^{i[t]}-e^{-i[t]}}{2i}~,\qquad
\qquad\operatorname{Cos}[t]\equiv\frac{e^{i[t]}+e^{-i[t]}}{2}~,
\end{equation}
so that
\begin{equation}
\Delta\operatorname{Sin}[t]=\operatorname{Cos}[t]~,\qquad\qquad\Delta
\operatorname{Cos}[t]=-\operatorname{Sin}[t]\ .
\end{equation}

As an illustration, consider phase-space rotations of the oscillator. \ The
umbral deformation of phase-space rotations,
\begin{equation}
\dot{x}=p,\quad\dot{p}=-x\qquad\longmapsto\qquad\Delta X(t)=P(t),\quad\Delta
P(t)=-X(t)\ ,
\end{equation}
readily yields, by directly deforming continuum solutions, the oscillatory
solutions,
\begin{equation}
X(t)=X(0)\operatorname{Cos}[t]+P(0)\operatorname{Sin}[t],\qquad
P(t)=P(0)\operatorname{Cos}[t]-X(0)\operatorname{Sin}[t]\ .
\label{UmbralOscillatorSolutions}%
\end{equation}
In view of (\ref{EisOrdinary}), and also%
\begin{equation}
(1+ia)=\sqrt{1+a^{2}}~e^{i\arctan(a)}\ ,
\end{equation}
the umbral sines and cosines in (\ref{UmbralOscillatorSolutions}) are seen to
amount to discrete phase-space spirals,
\begin{equation}
X(t)=(1+a^{2})^{\frac{t}{2a}}\Bigl (X(0)\cos(\omega t)+P(0)\sin(\omega
t)\Bigr )\ ,\quad P(t)=(1+a^{2})^{\frac{t}{2a}}\Bigl (P(0)\cos(\omega
t)-X(0)\sin(\omega t)\Bigr )\ ,
\end{equation}
with a frequency \emph{decreased} from the continuum value (i.e. 1) to
\begin{equation}
\omega=\arctan(a)/a\leq1~.
\end{equation}
So the frequency has become, effectively, the inverse of the cardinal tangent
function.\footnote{That is, for $\Theta\equiv\arctan(a)$, the spacing of the
zeros, period, etc, are scaled up by a factor of $\tanc(\Theta)\equiv
\frac{\tan(\Theta)}{\Theta}\geq1~.$ \ For complete periodicity on the time
lattice, one further needs return to the origin in an integral number of $N$
steps, thus a solution of $N=2\pi n/\arctan a$. Example: ~ For $a=1$, the
solutions' radius spirals out as $2^{t/2}$, while $\omega=\pi/4$, and the
period is $\tau=8$.} \ Note that the umbrally conserved quantity is,
\begin{equation}
2\mathcal{E}=X(0)^{2}+P(0)^{2}=(1+a^{2})^{\frac{-t}{a}}\Bigl (X(t)^{2}%
+P(t)^{2}\Bigr )\ ,
\end{equation}
such that $\Delta\mathcal{E}=0$, with the proper energy as the continuum limit.

\section{\textbf{Reduction from second-order differences to single term
recursions}}

In this section and the following, to conform to prevalent conventions, the
umbral variable will be denoted by $x$, instead of $t$. \ In this case there
is a natural way to think of the umbral correspondence that draws on familiar
quantum mechanics language \cite{rota}: \ The discrete difference equations
begin as operator statements, for operator $x$s and $T$s, but are then reduced
to equations involving classical-valued functions just by taking the matrix
element $\left\langle x\right\vert \cdots\left\vert vac\right\rangle $ where
$\left\vert vac\right\rangle $ is translationally invariant. \ The overall
$x$-\textbf{in}dependent non-zero constant $\left\langle x|vac\right\rangle $
is then ignored.

To be specific, consider Whittaker's equation
\href{http://www.convertit.com/Go/ConvertIt/Reference/AMS55.ASP?Res=150&Page=505}{(A\&S
13.1.31)} for $\mu=1/2$,
\begin{equation}
\left(  \partial_{x}^{2}+\frac{\kappa}{x}-\frac{1}{4}\right)  y(x)=0\ .
\label{Whittaker1/2}%
\end{equation}
This umbrally maps to the operator statement
\begin{equation}
\left(  \Delta^{2}+T~\frac{\kappa}{x}-\frac{1}{4}\right)  y(xT^{-1})=0\ .
\end{equation}
Considering either $y(xT^{-1})\cdot1\equiv Y(x)$, or else $\left\langle
x\right\vert y(xT^{-1})\left\vert vac\right\rangle =Y(x)\ \left\langle
x|vac\right\rangle $, this operator statement reduces to a classical
difference equation,%
\begin{equation}
Y(x+2a)-2Y(x+a)+Y(x)+\frac{\kappa a^{2}}{x+a}~Y(x+a)-\frac{a^{2}}{4}~Y(x)=0\ .
\label{UmbralWhittaker1/2}%
\end{equation}

Before using umbral mapping to convert continuous solutions of
(\ref{Whittaker1/2}) into discrete solutions \cite{LSNdO,LSNdOS} of
(\ref{UmbralWhittaker1/2}), here we note a simplification of the latter
equation upon choosing $a=2$, which amounts to setting the scale of $x$.
\ With this choice (\ref{UmbralWhittaker1/2}) collapses to a mere one-term
recursion. \ Shifting $x\rightarrow x-2$ this is
\begin{equation}
Y(x+2)=2\left(  \frac{x-2\kappa}{x}\right)  Y(x)\ . \label{UmCoulomb1stOrder}%
\end{equation}
Despite being a first-order difference equation, however, the solutions of
this equation still involve two independent \textquotedblleft constants of
summation\textquotedblright\ even for $x$ restricted to only integer values,
because the choice $a=2$ has decoupled adjacent vertical strips of unit width
on the complex $x$ plane. \ To be explicit, for integer $x>0$, forward
iteration gives \cite{benderorszag}%
\begin{equation}
Y(2k+1)=2^{k}\left(
%TCIMACRO{\dprod \limits_{j=1}^{k}}%
%BeginExpansion
{\displaystyle\prod\limits_{j=1}^{k}}
%EndExpansion
\frac{j-2\kappa}{j}\right)  Y\left(  1\right)  \ \ \ \text{and\ }%
\ \ Y(2k+2)=2^{k}\left(
%TCIMACRO{\dprod \limits_{j=1}^{k}}%
%BeginExpansion
{\displaystyle\prod\limits_{j=1}^{k}}
%EndExpansion
\frac{j-\kappa}{j}\right)  Y\left(  2\right)  \ ,\ \ \ \text{for integer
}k\geq0\text{ ,} \label{ProductRep}%
\end{equation}
with $Y\left(  1\right)  $ and $Y\left(  2\right)  $ the two independent
constants that determine values of $Y$ for all \emph{larger} odd and even
integer points, respectively.

Or, if generic $x$ is contemplated, the equation (\ref{UmCoulomb1stOrder}) has
elementary solutions, for arbitrary complex constants $C_{1}$ and $C_{2}$,
given by%
\begin{align}
Y(x)  &  =\frac{2^{x/2}\Gamma\left(  \frac{x}{2}-\kappa\right)  }%
{\Gamma\left(  \frac{x}{2}\right)  }~C_{1}+\frac{\left(  -2\right)  ^{x/2}%
}{\Gamma\left(  \frac{x}{2}\right)  \Gamma\left(  1-\frac{x}{2}+\kappa\right)
}~C_{2}\label{C1C2}\\
&  =\frac{2^{x/2}\Gamma\left(  \frac{x}{2}-\kappa\right)  }{\Gamma\left(
\frac{x}{2}\right)  }~\left(  C_{1}+\frac{1}{\pi}\left(  -1\right)
^{x/2}C_{2}\sin\pi\left(  \frac{x}{2}-\kappa\right)  \right)  \ .
\end{align}
In the second expression, we have used $\Gamma\left(  z\right)  \Gamma\left(
1-z\right)  =\pi/\sin\pi z$. \ Note the $C_{2}$ part of this elementary
solution differs from the $C_{1}$ part just through multiplication by a
particular complex function with period $2$. \ This is typical of solutions to
difference equations since \emph{any} such periodic factors are transparent to
$\Delta$, as mentioned in an earlier footnote \cite{ltw}. \ 

As expected, even for generic $x$ the constants $C_{1}$ and $C_{2}$\ may be
determined given $Y\left(  x\right)  $ at two judiciously chosen points, not
necessarily differing by an integer. \ For example, if $0<\kappa<1$,
\begin{equation}
C_{1}=\frac{\Gamma\left(  1+\kappa\right)  }{2^{1+\kappa}}~Y(2+2\kappa
)\ ,\ \ \ C_{2}=\frac{\pi}{\sin\pi\kappa}~C_{1}-\frac{1}{2}~\Gamma\left(
\kappa\right)  ~Y(2)\ .
\end{equation}
Moreover, poles and zeros of the solution are manifest either from the
$\Gamma$ functions in (\ref{C1C2}), or else from continued product
representations such as (\ref{ProductRep}). \ For the latter, either forward
or backward iterations of the first-order difference equation
(\ref{UmCoulomb1stOrder}) may be used. \ Schematically,%
\begin{equation}
Y(x)=\frac{\left(  2x-4-4\kappa\right)  \left(  2x-8-4\kappa\right)  \left(
2x-12-4\kappa\right)  \cdots}{\left(  x-2\right)  \left(  x-4\right)  \left(
x-6\right)  \cdots}\ ,
\end{equation}
or alternatively,%
\begin{equation}
Y(x)=\frac{x(x+2)(x+4)\cdots}{(2x-4\kappa)(2x+4(1-\kappa))(2x+4(2-\kappa
))\cdots}\ .
\end{equation}

Although both terms in (\ref{C1C2}) have zeroes, the $C_{1}$ term also has
poles while the $C_{2}$\ term has none --- it is an entire function of $x$ ---
and it is complex for any nonzero choice of $C_{2}$. \ Of course, since the
equation (\ref{UmCoulomb1stOrder}) is linear, real and imaginary parts may be
taken as separate real solutions. \ All this is evident in the following plots
for various selected integer $\kappa$.
%TCIMACRO{\FRAME{dtbpFU}{5.3195in}{3.544in}{0pt}{\Qcb{$\frac{2^{x/2}%
%\Gamma\left(  \frac{1}{2}x-\kappa\right)  }{\Gamma\left(  \frac{1}{2}x\right)
%}$ for $\kappa=1$, $2$, and $3$ in red, blue, and green.}}{}%
%{umbralcoulomb.eps}{\special{ language "Scientific Word";  type "GRAPHIC";
%width 5.3195in;  height 3.544in;  depth 0pt;  original-width 4.4832in;
%original-height 2.9776in;  cropleft "0";  croptop "1";  cropright "1";
%cropbottom "0";  filename 'UmbralCoulomb.eps';file-properties "XNPEU";}} }%
%BeginExpansion
\begin{center}
\includegraphics[
height=3.544in,
width=5.3195in
]%
{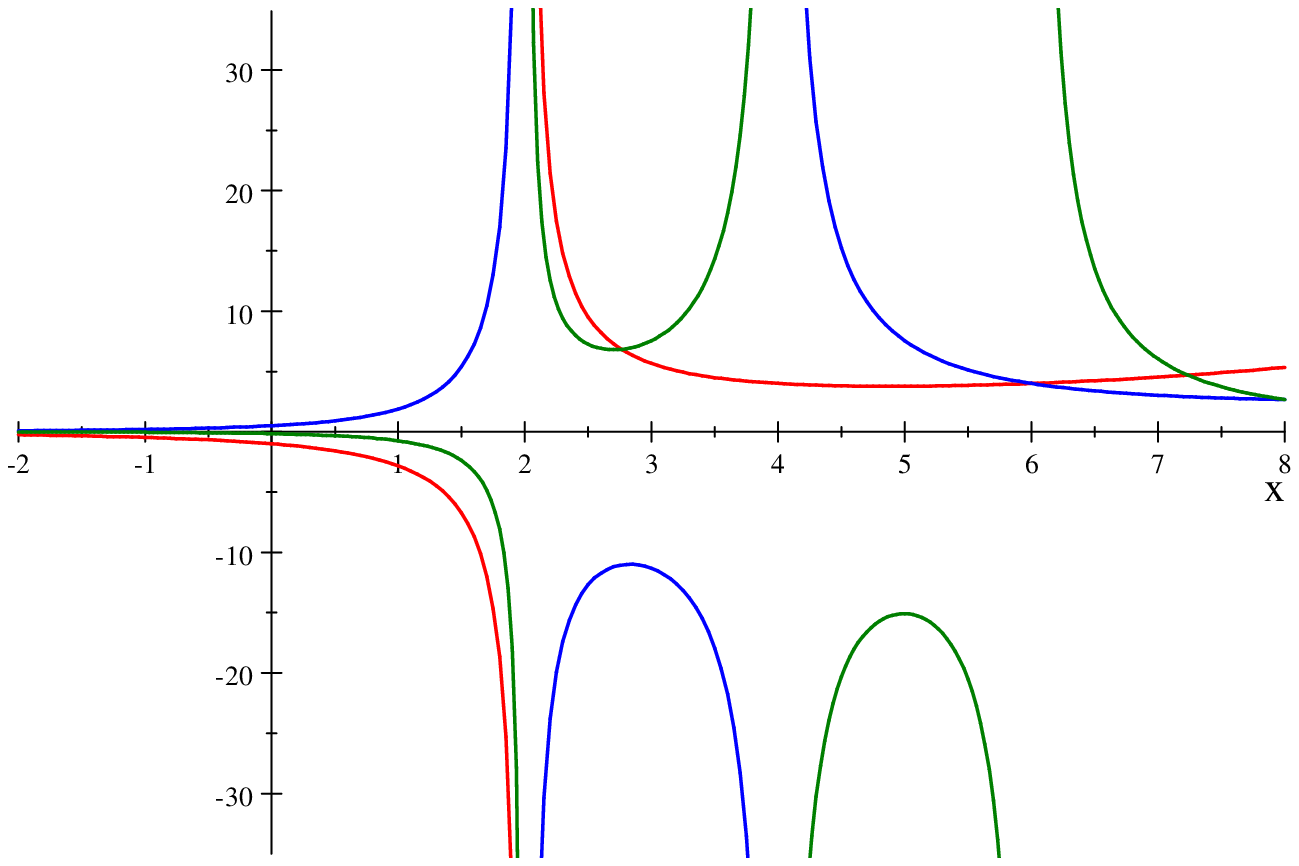}%
\\
$\frac{2^{x/2}\Gamma\left(  \frac{1}{2}x-\kappa\right)  }{\Gamma\left(
\frac{1}{2}x\right)  }$ for $\kappa=1$, $2$, and $3$ in red, blue, and green.
\end{center}
%EndExpansion%
%TCIMACRO{\FRAME{dtbpFU}{5.3195in}{3.5449in}{0pt}{\Qcb{$\frac{2^{x/2}\cos
%\pi\left(  \frac{1}{2}x\right)  }{\Gamma\left(  \frac{1}{2}x\right)
%\Gamma\left(  1-\frac{1}{2}x+\kappa\right)  }$ for $\kappa=1$, $2$, and $3$ in
%red, blue, and green.}}{}{umbralcoulombreal.eps}%
%width 5.3195in;  height 3.5449in;  depth 0pt;  original-width 4.4832in;
%original-height 2.9776in;  cropleft "0";  croptop "1";  cropright "1";
%cropbottom "0";  filename 'UmbralCoulombReal.eps';file-properties "XNPEU";}}
%}%
%BeginExpansion
\begin{center}
\includegraphics[
height=3.5449in,
width=5.3195in
]%
{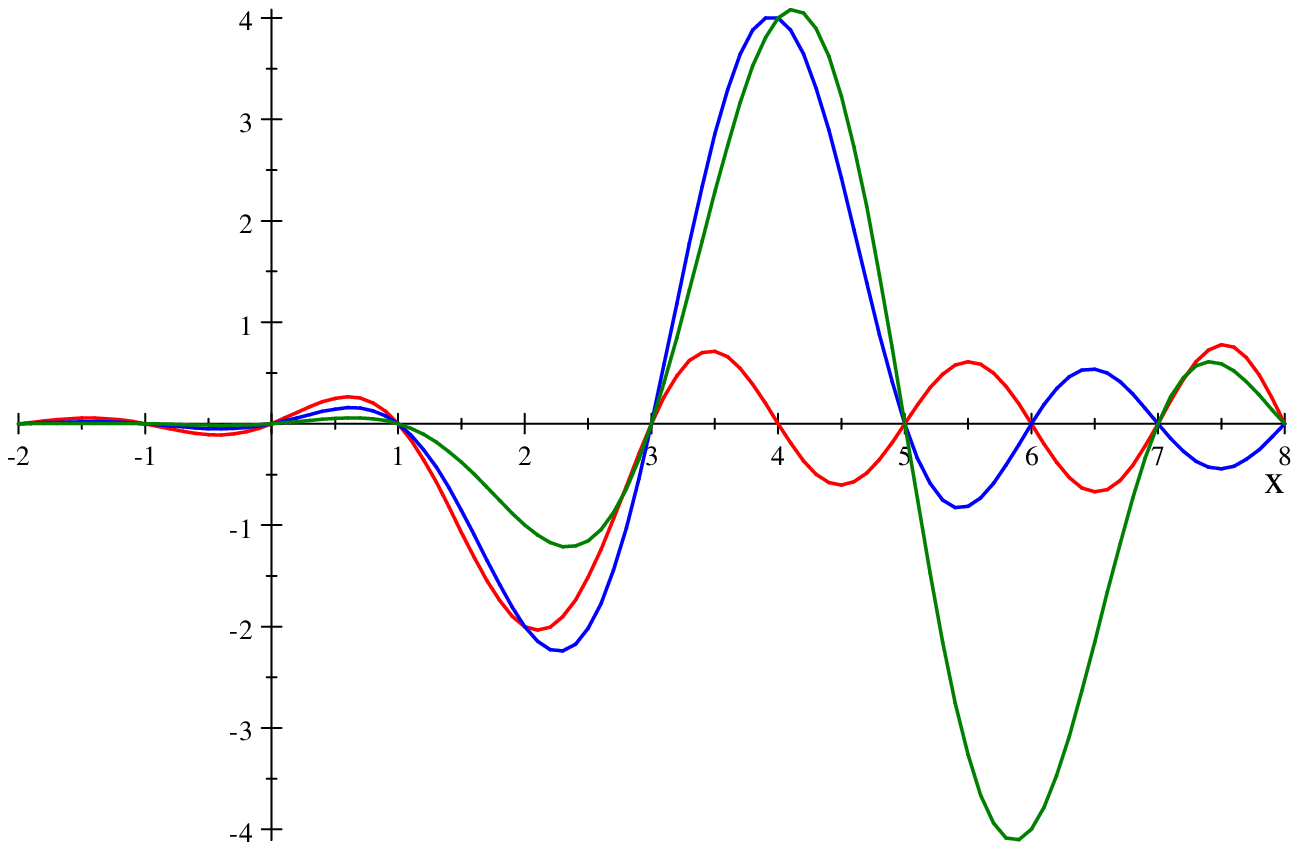}%
\\
$\frac{2^{x/2}\cos\pi\left(  \frac{1}{2}x\right)  }{\Gamma\left(  \frac{1}%
{2}x\right)  \Gamma\left(  1-\frac{1}{2}x+\kappa\right)  }$ for $\kappa=1$,
$2$, and $3$ in red, blue, and green.
\end{center}
%EndExpansion%
%TCIMACRO{\FRAME{dtbpFU}{5.3195in}{3.544in}{0pt}{\Qcb{$\frac{2^{x/2}\sin
%\pi\left(  \frac{1}{2}x\right)  }{\Gamma\left(  \frac{1}{2}x\right)
%\Gamma\left(  1-\frac{1}{2}x+\kappa\right)  }$ for $\kappa=1$, $2$, and $3$ in
%red, blue, and green.}}{}{umbralcoulombimag.eps}%
%width 5.3195in;  height 3.544in;  depth 0pt;  original-width 4.4832in;
%original-height 2.9776in;  cropleft "0";  croptop "1";  cropright "1";
%cropbottom "0";  filename 'UmbralCoulombImag.eps';file-properties "XNPEU";}}
%}%
%BeginExpansion
\begin{center}
\includegraphics[
height=3.544in,
width=5.3195in
]%
{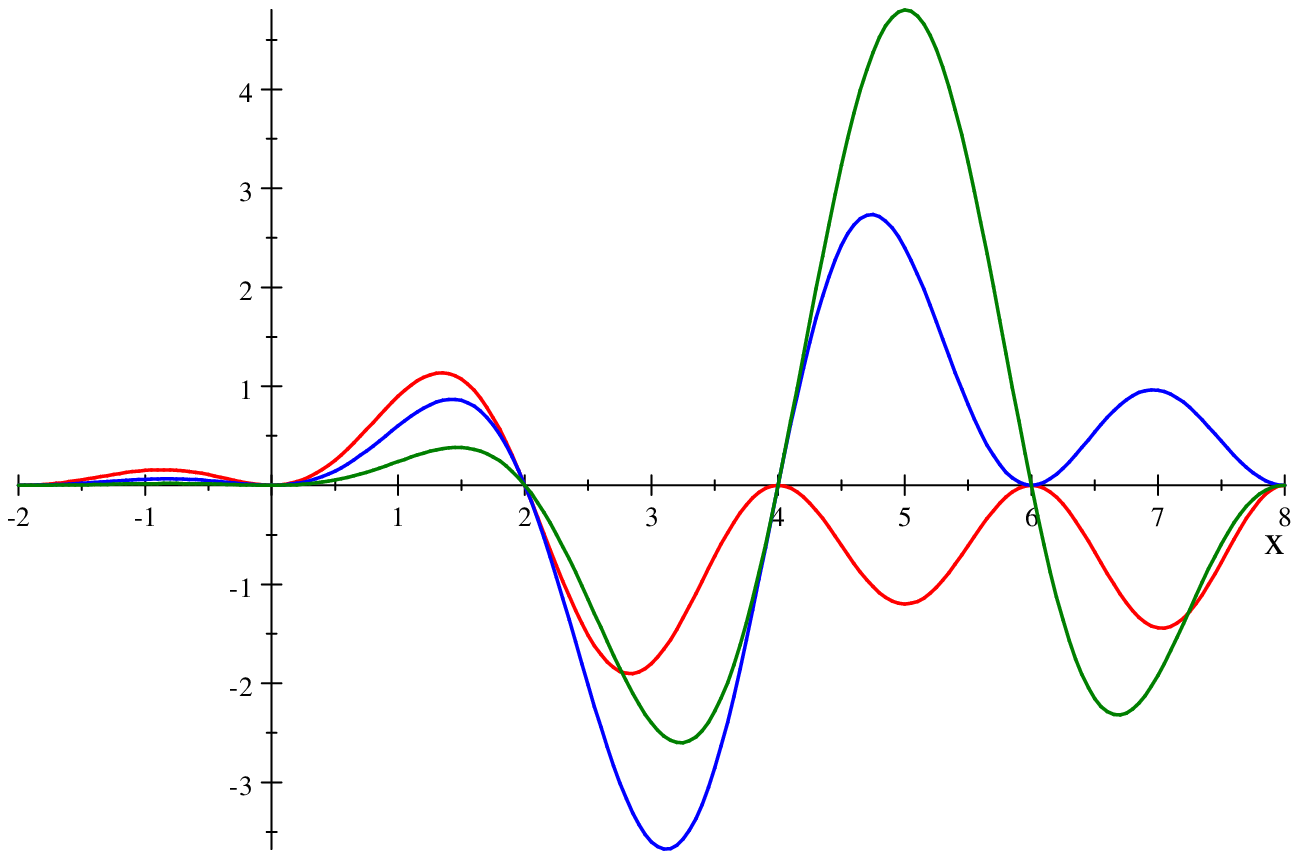}%
\\
$\frac{2^{x/2}\sin\pi\left(  \frac{1}{2}x\right)  }{\Gamma\left(  \frac{1}%
{2}x\right)  \Gamma\left(  1-\frac{1}{2}x+\kappa\right)  }$ for $\kappa=1$,
$2$, and $3$ in red, blue, and green.
\end{center}
%EndExpansion

\newpage

Collapse to a mere one-term recursion also occurs for an inverse-square
potential,
\begin{equation}
\left(  \partial_{x}^{2}+{\frac{\kappa}{x^{2}}}-\mu\right)  y(x)=0\ .
\label{InverseSquare}%
\end{equation}
For $\mu a^{2}=1$, which amounts to setting the scale of the \emph{energy} of
the solution, the umbral version of this equation reduces to
\begin{equation}
Y(x)=\frac{1}{2}\left(  1+\frac{\kappa a^{2}}{x(x+a)}\right)  Y(x+a)=\frac
{1}{2}\left(  1+\frac{a\kappa}{x}-\frac{a\kappa}{a+x}\right)  Y(x+a)\ .
\end{equation}
That is to say,%
\begin{equation}
Y(x+a)=\frac{2\left(  1+\frac{x}{a}\right)  \frac{x}{a}}{\left(  \frac{x}%
{a}+\frac{1+\sqrt{1-4\kappa}}{2}\right)  \left(  \frac{x}{a}+\frac
{1-\sqrt{1-4\kappa}}{2}\right)  }~Y(x)\ . \label{UmInverseSquare1stOrder}%
\end{equation}
Elementary solutions for generic $x$, for arbitrary \emph{complex} constants
$C_{1}$ and $C_{2}$, are given by%
\begin{align}
Y\left(  x\right)   &  =\frac{2^{x/a}}{\Gamma\left(  \frac{x}{a}+\frac
{1+\sqrt{1-4\kappa}}{2}\right)  \Gamma\left(  \frac{x}{a}+\frac{1-\sqrt
{1-4\kappa}}{2}\right)  }\left(  \Gamma\left(  1+\frac{x}{a}\right)
\Gamma\left(  \frac{x}{a}\right)  C_{1}+\frac{1}{\Gamma\left(  -\frac{x}%
{a}\right)  \Gamma\left(  1-\frac{x}{a}\right)  }~C_{2}\right) \\
&  =\frac{2^{x/a}\Gamma\left(  1+\frac{x}{a}\right)  \Gamma\left(  \frac{x}%
{a}\right)  }{\Gamma\left(  \frac{x}{a}+\frac{1+\sqrt{1-4\kappa}}{2}\right)
\Gamma\left(  \frac{x}{a}+\frac{1-\sqrt{1-4\kappa}}{2}\right)  }\left(
C_{1}-\frac{1}{\pi^{2}}~C_{2}\sin^{2}\left(  \frac{\pi x}{a}\right)  \right)
\ .
\end{align}
Again, the $C_{2}$ part of this elementary solution differs from the $C_{1}$
part just through multiplication by a particular complex function with period
$a$. \ And again, poles and zeros of these and other solutions are manifest
either from those of the $\Gamma$ functions, or else from a continued product
form, e.g.
\begin{equation}
Y(x)=\frac{(x^{2}+xa+\kappa a^{2})~((x+a)^{2}+(x+a)a+\kappa a^{2})\cdots
}{(2x(x+a))~(2(x+a)(x+2a))\cdots}\ .
\end{equation}

It is not surprising that (\ref{Whittaker1/2}) and (\ref{InverseSquare}) share
the privilege to become only first-order difference equations for specific
choices of $a$, as in (\ref{UmCoulomb1stOrder})\ and
(\ref{UmInverseSquare1stOrder}), because they are both special cases of
Whittaker's differential equation, as discussed in the next section. \ No
other linear second-order ODEs lead to umbral equations with this property.

\section{Discretization through hypergeometric recursion}

In this section we discuss several examples using umbral transform methods to
convert solutions of continuum differential equations directly into solutions
of the corresponding discretized equations. \ We use both Fourier and power
series umbral transforms.

As an explicit illustration of the umbral transform functional (\ref{neato}),
inserting the Fourier representation of the Airy function
\href{http://www.convertit.com/Go/ConvertIt/Reference/AMS55.ASP?Res=150&Page=447}{(A\&S
10.4.32)} yields
\begin{equation}
\operatorname{AiryAi}\left(  x\right)  \qquad\longmapsto\qquad
\operatorname*{UmAiryAi}\left(  x,a\right)  \equiv\operatorname{Re}\left(
\frac{1}{\pi}\int_{0}^{+\infty}e^{\frac{1}{3}ik^{3}}\left(  1+ika\right)
^{\frac{x}{a}}dk\right)  \ .
\end{equation}
This integral is expressed in terms of hypergeometric functions and
evaluated\ numerically in \href{#Appendix A}{Appendix A}.

Likewise, gaussians also map to hypergeometric functions, as may be obtained
by formal series manipulations:
\begin{gather}
e^{-x^{2}}\qquad\longmapsto\qquad G\left(  x,a\right)  \equiv\sum
_{n=0}^{\infty}\frac{(-)^{n}[x]^{2n}}{n!}=\sum_{n=0}^{\infty}\frac{1}%
{n!}\left(  -1\right)  ^{n}a^{2n}\frac{\Gamma\left(  \frac{x}{a}+1\right)
}{\Gamma\left(  \frac{x}{a}-2n+1\right)  }\\
=\sum_{n=0}^{\infty}\frac{\Gamma\left(  n-\frac{1}{2}\frac{x}{a}\right)
}{\Gamma\left(  -\frac{1}{2}\frac{x}{a}\right)  }\frac{\Gamma\left(
n+\frac{1}{2}-\frac{1}{2}\frac{x}{a}\right)  }{\Gamma\left(  \frac{1}{2}%
-\frac{1}{2}\frac{x}{a}\right)  }\frac{\left(  -4a^{2}\right)  ^{n}}%
{n!}\label{UmbralGaussSeries}\\
\equiv\left.  _{2}F_{0}\right.  \left(  -\frac{1}{2}\frac{x}{a},\frac{1}%
{2}\left(  1-\frac{x}{a}\right)  ;-4a^{2}\right)  \ , \label{UmbralGaussHyper}%
\end{gather}
where the
\href{http://www.convertit.com/Go/ConvertIt/Reference/AMS55.ASP?Res=150&Page=256}{reflection
and duplication formulas} were used to write
\begin{equation}
\frac{\Gamma\left(  \frac{x}{a}+1\right)  }{\Gamma\left(  \frac{x}%
{a}-2n+1\right)  }=\frac{4^{n}~\Gamma\left(  n-\frac{1}{2}\frac{x}{a}\right)
\Gamma\left(  n+\frac{1}{2}-\frac{1}{2}\frac{x}{a}\right)  }{\Gamma\left(
-\frac{1}{2}\frac{x}{a}\right)  \Gamma\left(  \frac{1}{2}-\frac{1}{2}\frac
{x}{a}\right)  }\ .
\end{equation}
While the series (\ref{UmbralGaussSeries}) actually has zero radius of
convergence, it is Borel summable, and the resulting regularized
hypergeometric function is well-defined.\ \ See \href{#Appendix B}{Appendix B}
for some related numerics.

For another example drawn from the familiar repertoire of continuum physics,
consider the confluent hypergeometric equation of Kummer
\href{http://www.convertit.com/Go/ConvertIt/Reference/AMS55.ASP?Res=150&Page=504}{(A\&S
13.1.1)}:%
\begin{equation}
x~y^{\prime\prime}+\left(  \beta-x\right)  ~y^{\prime}-\alpha~y=0\ ,
\label{KummerODE}%
\end{equation}
whose regular solution at $x=0$,
\href{http://mathworld.wolfram.com/ConfluentHypergeometricDifferentialEquation.html}{expressed
in various dialects}, is
\begin{equation}
y=\left.  _{1}F_{1}\right.  \left(  \alpha;\beta;x\right)  =M\left(
\alpha,\beta,x\right)  =\operatorname{KummerM}\left(  \alpha,\beta,x\right)
\ ,
\end{equation}
with series and integral representations%
\begin{align}
\left.  _{1}F_{1}\right.  \left(  \alpha;\beta;x\right)   &  =\sum
_{n=0}^{\infty}\frac{\Gamma\left(  \alpha+n\right)  }{\Gamma\left(
\alpha\right)  }\frac{\Gamma\left(  \beta\right)  }{\Gamma\left(
\beta+n\right)  }\frac{x^{n}}{n!}\label{1F1}\\
&  =\frac{\Gamma\left(  \beta\right)  }{\Gamma\left(  \alpha\right)
\Gamma\left(  \beta-\alpha\right)  }\int_{0}^{1}e^{xs}s^{\alpha-1}\left(
1-s\right)  ^{\beta-\alpha-1}ds\nonumber\\
&  =1+\frac{\alpha}{\beta}~x+\frac{1}{2}\frac{\alpha\left(  \alpha+1\right)
}{\beta\left(  \beta+1\right)  }~x^{2}+\frac{1}{6}\frac{\alpha\left(
\alpha+1\right)  \left(  \alpha+2\right)  }{\beta\left(  \beta+1\right)
\left(  \beta+2\right)  }~x^{3}+O\left(  x^{4}\right)  \ .\nonumber
\end{align}
The second, independent solution of (\ref{KummerODE}), with branch point at
$x=0$, is given by Tricomi's confluent hypergeometric function
\href{http://www.convertit.com/Go/ConvertIt/Reference/AMS55.ASP?Res=150&Page=504}{(A\&S
13.1.3)}, sometimes known as
\href{http://functions.wolfram.com/HypergeometricFunctions/HypergeometricU/}{HypergeometricU}%
:%
\begin{equation}
U\left(  \alpha,\beta,x\right)  =\frac{\pi}{\sin\pi\beta}\left(
\frac{M\left(  \alpha,\beta,x\right)  }{\Gamma\left(  1+\alpha-\beta\right)
\Gamma\left(  \beta\right)  }-x^{1-\beta}\frac{M\left(  1+\alpha-\beta
,2-\beta,x\right)  }{\Gamma\left(  \alpha\right)  \Gamma\left(  2-\beta
\right)  }\right)  \ . \label{HypergeometricU}%
\end{equation}

Invoking the umbral calculus for $x$, either of these confluent hypergeometric
functions can be mapped onto their umbral counterparts using%
\begin{equation}
\left.  _{1}F_{1}\right.  \left(  \alpha;\beta;x\right)  \qquad\longmapsto
\qquad\left.  _{2}F_{1}\right.  \left(  \alpha,-\frac{x}{a};\beta;-a\right)
\ ,
\end{equation}
where $\left.  _{2}F_{1}\right.  $ is the well-known Gauss hypergeometric
function
\href{http://www.convertit.com/Go/ConvertIt/Reference/AMS55.ASP?Res=150&Page=556}{(A\&S
15.1.1)}. \ This map from $\left.  _{1}F_{1}\right.  $ to $\left.  _{2}%
F_{1}\right.  $ follows from the basic monomial umbral map,
\begin{equation}
x^{n}\longmapsto\lbrack x]^{n}\equiv\left(  xT^{-1}\right)  ^{n}\cdot
1=a^{n}\frac{\Gamma\left(  \frac{x}{a}+1\right)  }{\Gamma\left(  \frac{x}%
{a}-n+1\right)  }=\left(  -a\right)  ^{n}\frac{\Gamma\left(  n-\frac{x}%
{a}\right)  }{\Gamma\left(  -\frac{x}{a}\right)  }\ ,
\end{equation}
and from the series (\ref{1F1}). \ When combined, these give the well-known
series representation of $\left.  _{2}F_{1}\right.  $.

Next, reconsider the one-dimensional Coulomb problem defined by Whittaker's
equation for general $\mu$
\href{http://www.convertit.com/Go/ConvertIt/Reference/AMS55.ASP?Res=150&Page=505}{(A\&S
13.1.31)}:
\begin{equation}
y^{\prime\prime}+\left(  -\frac{1}{4}+\frac{\kappa}{x}+\frac{\left(  \frac
{1}{4}-\mu^{2}\right)  }{x^{2}}\right)  y=0\ . \label{WhittakerODE}%
\end{equation}
Since $\kappa$ and $\mu$ are both arbitrary, this also encompasses the
inverse-square potential, (\ref{InverseSquare}).
\ \href{http://mathworld.wolfram.com/WhittakerFunction.html}{Exact solutions}
of this differential equation are
\begin{equation}
y\left(  x\right)  =C_{1}\operatorname{whittakerM}\left(  \kappa,\mu,x\right)
+C_{2}\operatorname{whittakerW}\left(  \kappa,\mu,x\right)  \ ,
\end{equation}%
\begin{equation}
\operatorname{whittakerM}\left(  \kappa,\mu,x\right)  =x^{\mu+1/2}%
e^{-x/2}\left.  _{1}F_{1}\right.  \left(  \mu-\kappa+\frac{1}{2}%
;2\mu+1;x\right)  \ ,
\end{equation}%
\begin{equation}
\operatorname{whittakerW}\left(  \kappa,\mu,x\right)  =x^{\mu+1/2}%
e^{-x/2}\left(
\begin{array}
[c]{c}%
\frac{\Gamma\left(  -2\mu\right)  }{\Gamma\left(  -\mu-\kappa+\frac{1}%
{2}\right)  }\left.  _{1}F_{1}\right.  \left(  \mu-\kappa+\frac{1}{2}%
;2\mu+1;x\right) \\
\\
+\frac{\Gamma\left(  2\mu\right)  }{\Gamma\left(  \mu-\kappa+\frac{1}%
{2}\right)  }~x^{-2\mu}\left.  _{1}F_{1}\right.  \left(  -\mu-\kappa+\frac
{1}{2};-2\mu+1;x\right)
\end{array}
\right)  \ .
\end{equation}
Umbral versions of these solutions are complicated by the exponential and
overall power factors in the classical relations between the $\left.
_{1}F_{1}\right.  $'s and the Whittaker functions, but this complication is
manageable. \ (In part this is because in the umbral calculus there are
\emph{no ordering ambiguities} \cite{ueno}.)

To obtain the umbral version of the Whittaker functions, we begin by
evaluating%
\begin{align}
e^{-\frac{1}{2}xT^{-1}}\left.  _{1}F_{1}\right.  \left(  \alpha;\beta
;xT^{-1}\right)  \cdot1  &  =\sum_{m=0}^{\infty}\sum_{n=0}^{\infty}\left(
-\frac{1}{2}\right)  ^{m}\frac{\frac{\Gamma\left(  \alpha+n\right)  }%
{\Gamma\left(  \alpha\right)  }}{\frac{\Gamma\left(  \beta+n\right)  }%
{\Gamma\left(  \beta\right)  }}\frac{\left[  x\right]  ^{m+n}}{m!n!}%
\nonumber\\
&  =\left(  1-\frac{a}{2}\right)  ^{\frac{x}{a}}\left.  _{2}F_{1}\right.
\left(  \alpha,-\frac{x}{a};\beta;\frac{2a}{a-2}\right)  \ ,
\label{ExpOnConfluent}%
\end{align}
where we have performed the sum over $m$ first, to obtain%
\begin{equation}
\sum_{m=0}^{\infty}\left(  -\frac{1}{2}\right)  ^{m}\frac{1}{\Gamma\left(
\frac{x}{a}-n-m+1\right)  }\frac{a^{m}}{m!}=\frac{1}{\Gamma\left(  \frac{x}%
{a}-n+1\right)  }\left(  1-\frac{a}{2}\right)  ^{\frac{x-na}{a}}\ .
\end{equation}
The sum over $n$ then gives the Gauss hypergeometric function in
(\ref{ExpOnConfluent}).

Next, to deal with the umbral deformations of the Whittaker functions, we need
to use the continuation of (\ref{PosIntPower}) and (\ref{NegIntPower}) to an
\emph{arbitrary} power of $xT^{-1}$, namely,%
\begin{equation}
\left(  xT^{-1}\right)  ^{\gamma}=a^{\gamma}\frac{\Gamma\left(  \frac{x}%
{a}+1\right)  }{\Gamma\left(  \frac{x}{a}-\gamma+1\right)  }T^{-\gamma}~.
\label{ArbPower}%
\end{equation}
This continuation leads to the following:
\begin{align}
\left(  xT^{-1}\right)  ^{\gamma}e^{-\frac{1}{2}xT^{-1}}\left.  _{1}%
F_{1}\right.  \left(  \alpha;\beta;xT^{-1}\right)  \cdot1  &  =a^{\gamma}%
\frac{\Gamma\left(  \frac{x}{a}+1\right)  }{\Gamma\left(  \frac{x}{a}%
-\gamma+1\right)  }T^{-\gamma}e^{-\frac{1}{2}xT^{-1}}\left.  _{1}F_{1}\right.
\left(  \alpha;\beta;xT^{-1}\right)  \cdot1\nonumber\\
&  =a^{\gamma}\frac{\Gamma\left(  \frac{x}{a}+1\right)  }{\Gamma\left(
\frac{x}{a}-\gamma+1\right)  }e^{-\frac{1}{2}\left(  x-\gamma a\right)
T^{-1}}\left.  _{1}F_{1}\right.  \left(  \alpha;\beta;\left(  x-\gamma
a\right)  T^{-1}\right)  \cdot1\ .
\end{align}
Thus we obtain the umbral map%
\begin{equation}
x^{\gamma}e^{-\frac{1}{2}x}\left.  _{1}F_{1}\right.  \left(  \alpha
;\beta;x\right)  \longmapsto\frac{\Gamma\left(  \frac{x}{a}+1\right)  }%
{\Gamma\left(  \frac{x}{a}-\gamma+1\right)  }~a^{\gamma}\left(  1-\frac{a}%
{2}\right)  ^{\frac{x}{a}-\gamma}\left.  _{2}F_{1}\right.  \left(
\alpha,\gamma-\frac{x}{a};\beta;\frac{2a}{a-2}\right)  \ .
\end{equation}
Finally then, specializing to the relevant $\alpha$, $\beta$, and $\gamma$, we
find the umbral Whittaker functions. \ In particular, \
\begin{equation}
\operatorname{whittakerM}\left(  \kappa,\mu,x\right)  \longmapsto\frac
{\Gamma\left(  \frac{x}{a}+1\right)  }{\Gamma\left(  \frac{x}{a}-\mu+\frac
{1}{2}\right)  }~a^{\mu+1/2}\left(  1-\frac{a}{2}\right)  ^{\frac{x}{a}%
-\mu-\frac{1}{2}}\left.  _{2}F_{1}\right.  \left(  \mu+\frac{1}{2}-\kappa
,\mu+\frac{1}{2}-\frac{x}{a};2\mu+1;\frac{2a}{a-2}\right)  \ .
\label{UmbralWhittaker}%
\end{equation}

This result for general $a$ exhibits what is special about the choice $a=2$,
as exploited in the previous section. \ To realize that choice from
(\ref{UmbralWhittaker}) requires taking a limit $a\nearrow2$, hence it
requires the asymptotic behavior of the Gauss hypergeometric function
\href{http://www.convertit.com/Go/ConvertIt/Reference/AMS55.ASP?Res=150&Page=559}{(A\&S
15.3.7)}:%
\begin{equation}
\left.  _{2}F_{1}\right.  \left(  \alpha,\beta;\gamma;z\right)
\underset{z\rightarrow-\infty}{\sim}\frac{\Gamma\left(  \gamma\right)
}{\Gamma\left(  \beta\right)  }\frac{\Gamma\left(  \beta-\alpha\right)
}{\Gamma\left(  \gamma-\alpha\right)  }\left(  -z\right)  ^{-\alpha}%
+\frac{\Gamma\left(  \gamma\right)  }{\Gamma\left(  \alpha\right)  }%
\frac{\Gamma\left(  \alpha-\beta\right)  }{\Gamma\left(  \gamma-\beta\right)
}\left(  -z\right)  ^{-\beta}\ . \label{AsympGauss}%
\end{equation}
Now with sufficient care, $a=2$ solutions can be coaxed from the umbral
version of $\operatorname{whittakerM}$ in (\ref{UmbralWhittaker}), and/or the
corresponding umbral counterpart of $\operatorname{whittakerW}$, upon taking
$\lim_{a\nearrow2}$ and making use of (\ref{AsympGauss}). \ Moreover, in
principle the umbral correspondents of both Whittaker functions could be used
to obtain from this limit a solution with two arbitrary constants. \ 

On the other hand, for $a=2$, the umbral equation corresponding to
(\ref{WhittakerODE}) again reduces to a one-term recursion, namely,%
\begin{equation}
Y(x+2)=\frac{2\left(  x+2\right)  \left(  x-2\kappa\right)  }{\left(
x+1+2\mu\right)  \left(  x+1-2\mu\right)  }~Y(x)\ .
\end{equation}
For generic $x$, solutions for arbitrary complex constants $C_{1}$ and $C_{2}$
are then given by%
\begin{align}
Y\left(  x\right)   &  =\frac{2^{x/2}}{\Gamma\left(  \frac{x}{2}+\frac{1}%
{2}+\mu\right)  \Gamma\left(  \frac{x}{2}+\frac{1}{2}-\mu\right)  }\left(
\Gamma\left(  1+\frac{x}{2}\right)  \Gamma\left(  \frac{x}{2}-\kappa\right)
C_{1}+\frac{1}{\Gamma\left(  -\frac{x}{2}\right)  \Gamma\left(  1+\kappa
-\frac{x}{2}\right)  }~C_{2}\right) \label{a=2UmbralWhittaker}\\
&  =\frac{2^{x/2}\Gamma\left(  1+\frac{x}{2}\right)  \Gamma\left(  \frac{x}%
{2}-\kappa\right)  }{\Gamma\left(  \frac{x}{2}+\frac{1}{2}+\mu\right)
\Gamma\left(  \frac{x}{2}+\frac{1}{2}-\mu\right)  }\left(  C_{1}+\frac{1}%
{\pi^{2}}~C_{2}\sin\left(  \frac{\pi x}{2}\right)  \sin\pi\left(  \frac{x}%
{2}-\kappa\right)  \right)  \ ,
\end{align}
which agrees with (\ref{C1C2}) when $\mu=1/2$, of course. \ As in that
previous special case, the $C_{2}$ part of (\ref{a=2UmbralWhittaker}) differs
from the $C_{1}$ part just through multiplication by a particular complex
function with period $2$ \cite{ltw}.

We graph some examples to show the differences between the Whittaker functions
and their umbral counterparts, for $a=1$.%
%TCIMACRO{\FRAME{dtbpFU}{5.3039in}{3.5293in}{0pt}%
%{\Qcb{$\operatorname{whittakerM}\left(  \kappa,1/2,x\right)  $ for $\kappa=1$,
%$2$, and $3$ in red, blue, and green.}}{}{whittaker.eps}%
%width 5.3039in;  height 3.5293in;  depth 0pt;  original-width 5.3056in;
%original-height 3.5215in;  cropleft "0";  croptop "1";  cropright "1";
%cropbottom "0";  filename 'Whittaker.eps';file-properties "XNPEU";}} }%
%BeginExpansion
\begin{center}
\includegraphics[
height=3.5293in,
width=5.3039in
]%
{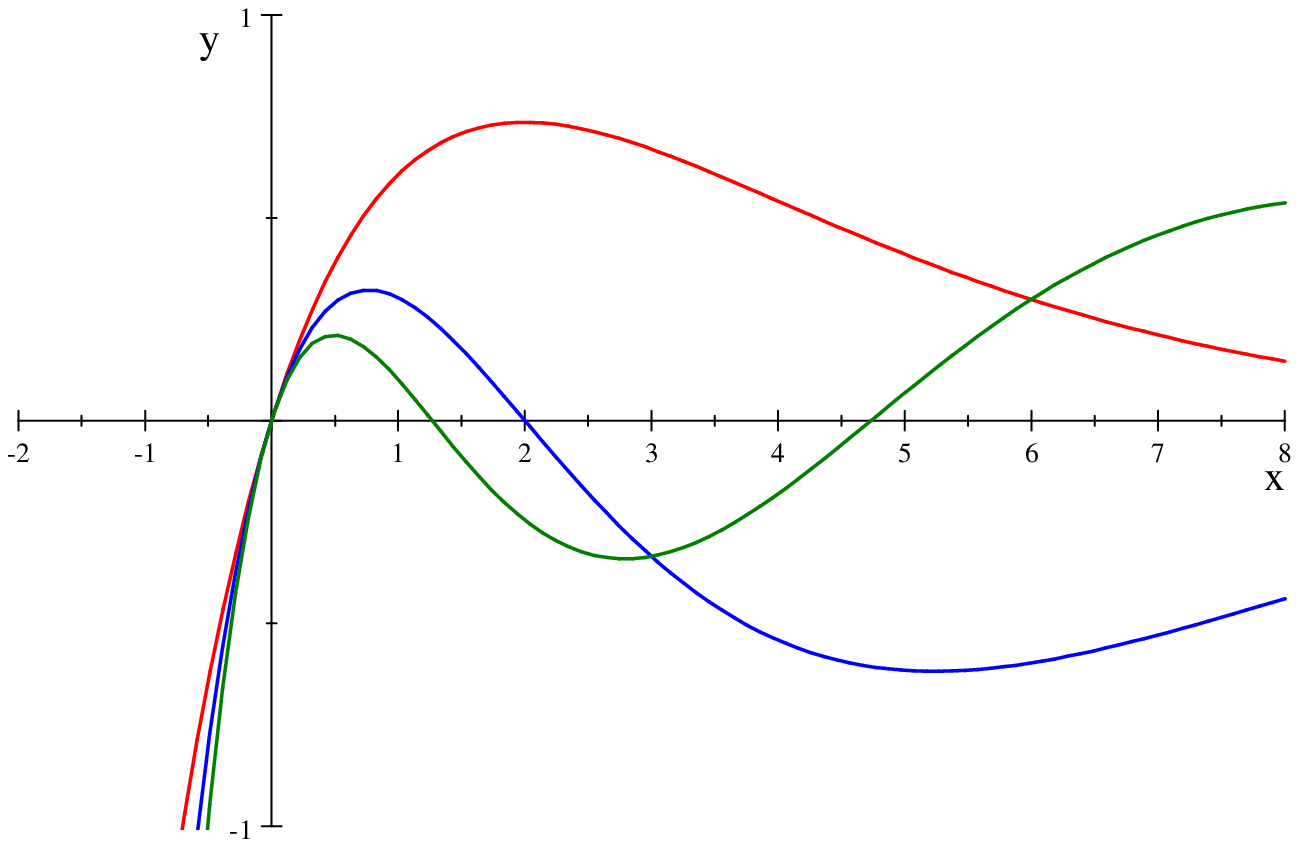}%
\\
$\operatorname{whittakerM}\left(  \kappa,1/2,x\right)  $ for $\kappa=1$, $2$,
and $3$ in red, blue, and green.
\end{center}
%EndExpansion
%TCIMACRO{\FRAME{dtbpFU}{5.3195in}{3.3996in}{0pt}{\Qcb{Umbral
%$\operatorname{whittakerM}\left(  \kappa,1/2,x\right)  $ for $a=1$, and for
%$\kappa=1$, $2$, and $3$ in red, blue, and green.}}{}{umbralwhittaker.eps}%
%width 5.3195in;  height 3.3996in;  depth 0pt;  original-width 5.2632in;
%original-height 3.3529in;  cropleft "0";  croptop "1";  cropright "1";
%cropbottom "0";  filename 'UmbralWhittaker.eps';file-properties "XNPEU";}} }%
%BeginExpansion
\begin{center}
\includegraphics[
height=3.3996in,
width=5.3195in
]%
{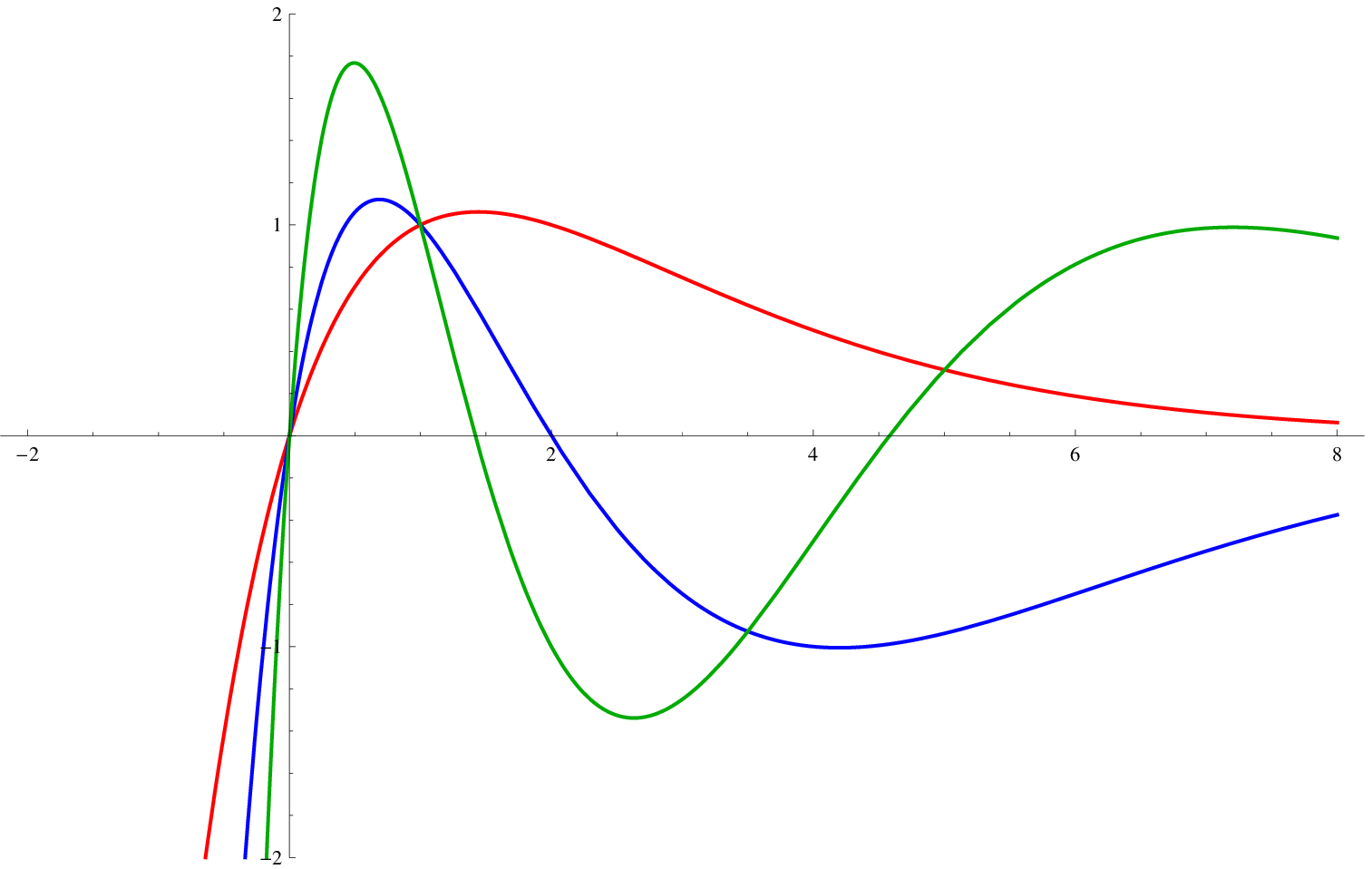}%
\\
Umbral $\operatorname{whittakerM}\left(  \kappa,1/2,x\right)  $ for $a=1$, and
for $\kappa=1$, $2$, and $3$ in red, blue, and green.
\end{center}
%EndExpansion

\newpage

The examples above are specific illustrations of combinatorics that may be
summarized in a few \emph{umbral hypergeometric mapping lemmata}, the simplest being

\fbox{\textbf{Lemma 1:}}
\begin{equation}
\left.  _{p}F_{q}\right.  (\alpha_{1},...,\alpha_{p};\beta_{1},...,\beta
_{q};x)~~~~~~\longmapsto~~~~~~\left.  _{p+1}F_{q}\right.  (\alpha
_{1},...,\alpha_{p},-x/a;\beta_{1},...,\beta_{q};-a)\ , \label{Hyper1}%
\end{equation}
where the series representation of the generalized hypergeometric function
$\left.  _{p}F_{q}\right.  $\ is\footnote{\noindent Recall results from using
the ratio test to determine the radius of convergence for the $\left.
_{p}F_{q}\right.  \left(  \alpha_{1},\cdots,\alpha_{p};\beta_{1},\cdots
,\beta_{q};x\right)  $ series:
\par
If $p<q+1$ then the ratio of coefficients tends to zero. \ This implies that
the series converges for any finite value of $x$.
\par
If $p=q+1$ then the ratio of coefficients tends to one, hence the series
converges for $|x|<1$ and diverges for $|x|>1$.
\par
If $p>q+1$ then the ratio of coefficients grows without bound. The series is
then divergent or asymptotic, and is a symbolic shorthand for the solution to
a differential equation.}
\begin{equation}
\left.  _{p}F_{q}\right.  \left(  \alpha_{1},\cdots,\alpha_{p};\beta
_{1},\cdots,\beta_{q};x\right)  =\frac{\Gamma\left(  \beta_{1}\right)
\cdots\Gamma\left(  \beta_{q}\right)  }{\Gamma\left(  \alpha_{1}\right)
\cdots\Gamma\left(  \alpha_{p}\right)  }\sum_{n=0}^{\infty}\frac{\Gamma\left(
\alpha_{1}+n\right)  \cdots\Gamma\left(  \alpha_{p}+n\right)  }{\Gamma\left(
\beta_{1}+n\right)  \cdots\Gamma\left(  \beta_{q}+n\right)  }\frac{x^{n}}%
{n!}\ .
\end{equation}
A proof of (\ref{Hyper1}) follows from formal manipulations of these series.

The umbral version of a more general class of functions is obtained by
replacing $x\rightarrow xT^{-1}$ in functions of $x^{k}$ for some fixed
positive integer $k$. \ Thus, again for hypergeometric functions, we have

\fbox{\textbf{Lemma 2:}}
\begin{gather}
\left.  _{p}F_{q}\right.  \left(  \alpha_{1},\cdots,\alpha_{p};\beta
_{1},\cdots,\beta_{q};x^{k}\right)  ~~~~~~\longmapsto\label{Hyper2}\\
\left.  _{p+k}F_{q}\right.  \left(  \alpha_{1},\cdots,\alpha_{p},\frac{1}%
{k}\left(  -\frac{x}{a}\right)  ,\frac{1}{k}\left(  1-\frac{x}{a}\right)
,\cdots,\frac{1}{k}\left(  k-1-\frac{x}{a}\right)  ;\beta_{1},\cdots,\beta
_{q};\left(  -ak\right)  ^{k}\right)  \ .\nonumber
\end{gather}
And again, a proof follows from formal series expansions.

Multiplication by exponentials produces only minor modifications of these
general results, as was discussed above in the context of Whittaker functions, namely,

\fbox{\textbf{Lemma 3:}}
\begin{gather}
e^{\lambda x}\left.  _{p}F_{q}\right.  \left(  \alpha_{1},\cdots,\alpha
_{p};\beta_{1},\cdots,\beta_{q};x^{k}\right)  ~~~~~~\longmapsto\label{Hyper3}%
\\
\left(  1+a\lambda\right)  ^{\frac{x}{a}}\left.  _{p+k}F_{q}\right.  \left(
\alpha_{1},\cdots,\alpha_{p},\frac{1}{k}\left(  -\frac{x}{a}\right)  ,\frac
{1}{k}\left(  1-\frac{x}{a}\right)  ,\cdots,\frac{1}{k}\left(  k-1-\frac{x}%
{a}\right)  ;\beta_{1},\cdots,\beta_{q};\left(  \frac{-ak}{1+a\lambda}\right)
^{k}\right)  \ .\nonumber
\end{gather}

In addition, multiplication by an overall power of $x$ gives

\fbox{\textbf{Lemma 4:}}
\begin{gather}
x^{\gamma}e^{\lambda x}\left.  _{p}F_{q}\right.  \left(  \alpha_{1}%
,\cdots,\alpha_{p};\beta_{1},\cdots,\beta_{q};x^{k}\right)  ~~~~~~\longmapsto
\\
\hspace{-0.3in}\frac{\Gamma\left(  \frac{x}{a}+1\right)  a^{\gamma}\left(
1+a\lambda\right)  ^{\frac{x}{a}-\gamma}}{\Gamma\left(  \frac{x}{a}%
-\gamma+1\right)  }\left.  _{p+k}F_{q}\right.  \left(  \alpha_{1}%
,\cdots,\alpha_{p},\frac{\gamma-\frac{x}{a}}{k},\frac{1+\gamma-\frac{x}{a}}%
{k},\cdots,\frac{k-1+\gamma-\frac{x}{a}}{k};\beta_{1},\cdots,\beta_{q};\left(
\frac{-ak}{1+a\lambda}\right)  ^{k}\right)  \ .\nonumber
\end{gather}

\section{Wave propagation}

Given the umbral features of discrete time and space equations discussed
above, separately, it is natural to combine the two.

For example, the umbral version of simple plane waves in 1+1 spacetime would
obey an equation of the type \cite{floreanini,ltw},
\begin{equation}
(\Delta_{x}^{2}-\Delta_{t}^{2})~F=0\ ,
\end{equation}
on a time-lattice with spacing $a$ and a space-lattice with spacing $b$, not
necessarily such that $b=a$ in all spacetime regions. \ For generic frequency,
wavenumber and velocity, the basic solutions are
\begin{equation}
f=e^{i(\omega t-kx)}\longmapsto F=\left(  1+i\omega a\right)  ^{t/a}\left(
1-ikb\right)  ^{x/b}\ .
\end{equation}
\ For right-moving waves, say, these have phase velocity
\begin{equation}
v(\omega,k)=\frac{\omega}{k}~\frac{a\arcsin(b)}{b\arcsin(a)}\ .
\end{equation}
Thus, the effective index of refraction in the discrete medium is
$(b\arcsin(a))/(a\arcsin(b))$, i.e. modified from 1. \ Small inhomogeneities
of $a$ and $b$ in the fabric of spacetime over large regions could therefore
yield interesting effects.

Technically, a more challenging application of umbral methods involves
nonlinear, solitonic phenomena \cite{ckz}, such as the one-soliton solution of
the continuum
\href{http://en.wikipedia.org/wiki/Sine-Gordon_equation}{Sine-Gordon
equation},
\begin{equation}
(\partial_{x}^{2}-\partial_{t}^{2})f\left(  x,t\right)  =\sin(f\left(
x,t\right)  )\ ,\ \ \ f_{\text{SG}}\left(  x,t\right)  =4\arctan\left(
me^{\frac{x-vt}{\sqrt{1-v^{2}}}}\right)  \ .
\end{equation}
The corresponding umbral deformation of the PDE itself would now also involve
a deformed potential $\sin(f(xT_{x}^{-1},tT_{t}^{-1}))\cdot1$. \ But rather
than tackling this difficult nonlinear difference equation, one may instead
use the umbral transform (\ref{neato}) to infer that $f_{\text{SG}}\left(
x,t\right)  $ maps to
\begin{equation}
F_{\text{SG}}\left(  a,b;x,t\right)  =\int_{-\infty}^{\infty}\!\!\frac{d\chi
d\tau d\omega dk}{\left(  2\pi\right)  ^{2}}~4\arctan\left(  me^{\frac
{\chi-v\tau}{\sqrt{1-v^{2}}}}\right)  ~e^{ik\chi-i\omega\tau}\left(  1+i\omega
a\right)  ^{t/a}\left(  1-ikb\right)  ^{x/b}\ .
\end{equation}
The \href{http://en.wikipedia.org/wiki/KdV}{continuum Korteweg--de Vries
soliton} is likewise mapped:
\begin{align}
f_{\text{KdV}}(x,t)  &  =\frac{v}{2}\operatorname{sech}^{2}(\frac{\sqrt{v}}%
{2}(x-vt))~\longmapsto~\\
F_{\text{KdV}}\left(  a,b;x,t\right)   &  =\int_{-\infty}^{\infty}%
\!\!\frac{d\chi d\tau d\omega dk}{\left(  2\pi\right)  ^{2}}~\frac{v}%
{2}~\operatorname{sech}^{2}\left(  \frac{\sqrt{v}}{2}(\chi-v\tau)\right)
~e^{ik\chi-i\omega\tau}\left(  1+i\omega a\right)  ^{t/a}\left(  1-ikb\right)
^{x/b}\ .\nonumber
\end{align}
Closed-form evaluations of these Fourier integrals are not available, but the
physical effects of the discretization could be investigated numerically, and
compared to the Lax pair integrability machinery of \cite{levi}, or to the
results on a variety of discrete KdVs in \cite{schiff}, or to other studies
\cite{Grammaticos}.

However, a more accessible example of umbral effects on solitons may be found
in the original \href{http://en.wikipedia.org/wiki/Toda_lattice}{Toda lattice
model} \cite{Toda}. \ For this model the spatial variable is already discrete,
usually with spacing $b=1$ so $x=n$ is an integer, while the time $t$ is
continuous. \ The equations of motion in that case are%
\begin{equation}
\frac{\partial q\left(  n,t\right)  }{\partial t}=p\left(  n,t\right)
\ ,\ \ \ \frac{\partial p\left(  n,t\right)  }{\partial t}=-\left(
e^{-\left(  q\left(  n+1,t\right)  -q\left(  n,t\right)  \right)
}-e^{-\left(  q\left(  n,t\right)  -q\left(  n-1,t\right)  \right)  }\right)
\ , \label{TodaEofM}%
\end{equation}
for integer $n$. \ Though $x=n$ is discrete, nevertheless there are exact
multi-soliton solutions valid for all continuous $t$, as is well-known.

Specific one-soliton Toda solutions are given for constant $\alpha$, $\beta$,
$\gamma$, and $q_{0}$ by
\begin{align}
q\left(  n,t\right)   &  =q_{0}+\log\left(  \frac{1+\alpha\exp\left(  -\beta
n+\gamma t\right)  }{1+\alpha\exp\left(  -\beta\left(  n+1\right)  +\gamma
t\right)  }\right)  \ ,\label{TodaSoliton}\\
p\left(  n,t\right)   &  =\alpha\gamma\left(  \frac{e^{-n\beta+\gamma t}%
}{\alpha e^{-n\beta+\gamma t}+1}-\frac{e^{-\left(  n+1\right)  \beta+\gamma
t}}{\alpha e^{-\left(  n+1\right)  \beta+\gamma t}+1}\right)  \ ,
\end{align}
\emph{provided} that%
\begin{equation}
\gamma=\pm2\sinh\left(  \frac{\beta}{2}\right)  \ . \label{SolutionCondition}%
\end{equation}
So the soliton's velocity is just $v=\pm\frac{2}{\beta}\sinh\left(
\frac{\beta}{2}\right)  $.

While obtained only for discrete $x=n$, for plotting purposes $q\left(
n,t\right)  $ may be interpolated for any $x$ (see graph below). \ To carry
out the complete umbral deformation of this system, it is then only necessary
to discretize $t$ in the equations of motion (\ref{TodaEofM}). \ Consider what
effects this approach to discrete time has on the specified one-soliton
solutions. \ 

To that end, expand the exact solutions in (\ref{TodaSoliton}) as series, \
\begin{equation}
q\left(  n,t\right)  =q_{0}+\sum_{k=1}^{\infty}\frac{\left(  -\alpha e^{-\beta
n}\right)  ^{k}}{k}\left(  e^{-k\beta}-1\right)  \exp\left(  \gamma kt\right)
\ .
\end{equation}
Upon umbralizing $t$, the one-soliton solutions then map as%
\begin{equation}
q\left(  n,t\right)  \longmapsto Q\left(  n,t\right)  \equiv q_{0}+\sum
_{k=1}^{\infty}\frac{\left(  -\alpha e^{-\beta n}\right)  ^{k}}{k}\left(
e^{-k\beta}-1\right)  (1+\gamma ka)^{t/a}\ , \label{UmbralTodaSoliton}%
\end{equation}
and these are guaranteed to give solutions to the umbral operator equations of
motion,%
\begin{align}
\Delta q\left(  n,tT^{-1}\right)   &  \equiv\frac{1}{a}\left(  T-1\right)
q\left(  n,tT^{-1}\right)  =p\left(  n,tT^{-1}\right)  \ ,\\
\Delta p\left(  n,tT^{-1}\right)   &  \equiv\frac{1}{a}\left(  T-1\right)
p\left(  n,tT^{-1}\right)  =-\left(  e^{-\left(  q\left(  n+1,tT^{-1}\right)
-q\left(  n,tT^{-1}\right)  \right)  }-e^{-\left(  q\left(  n,tT^{-1}\right)
-q\left(  n-1,tT^{-1}\right)  \right)  }\right)  \ ,
\end{align}
upon projecting onto a translationally invariant \textquotedblleft
vacuum\textquotedblright\ (i.e. $Q\left(  n,t\right)  \equiv q\left(
n,tT^{-1}\right)  \cdot1$).

Now, for integer time steps, $t/a=m$, consider the series at hand:%
\begin{equation}
S\left(  m,c,z\right)  =\sum_{k=1}^{\infty}\frac{z^{k}}{k}\left(  e^{-k\beta
}-1\right)  (1+ck)^{m}=\ln\left(  \frac{1-z}{1-ze^{-\beta}}\right)
+\sum_{j=1}^{m}c^{j}\binom{m}{j}R\left(  j,z\right)  \ , \label{TodaSeries}%
\end{equation}
where $c=\gamma a$, $z=-\alpha e^{-\beta n}$, and where for $j>0$,
\begin{equation}
R\left(  j,z\right)  =\sum_{k=0}^{\infty}\left(  e^{-k\beta}-1\right)
z^{k}k^{j-1}\equiv\Phi\left(  e^{-\beta}z,1-j,0\right)  -\Phi\left(
z,1-j,0\right)  \ .
\end{equation}
Fortunately, for positive integer $t/a$, we only need the
\href{http://mathworld.wolfram.com/LerchTranscendent.html}{Lerch transcendent
function},
\begin{equation}
\Phi\left(  z,s,r\right)  =\sum_{k=0}^{\infty}\frac{z^{k}}{\left(  r+k\right)
^{s}}\ , \label{Lerch}%
\end{equation}
for those cases where the sums are expressible as elementary functions. \ For
example,%
\begin{equation}
\sum_{k=0}^{\infty}z^{k}=\frac{1}{1-z}\ ,\ \ \ \sum_{k=0}^{\infty}z^{k}%
k=\frac{z}{\left(  1-z\right)  ^{2}}\ ,\ \ \ \sum_{k=0}^{\infty}z^{k}%
k^{2}=\frac{z+z^{2}}{\left(  1-z\right)  ^{3}}\ ,\ \ \ \sum_{k=0}^{\infty
}z^{k}k^{3}=\frac{z+4z^{2}+z^{3}}{\left(  1-z\right)  ^{4}}\ .
\end{equation}
The $\ln\left(  \cdots\right)  $ term on the RHS of (\ref{TodaSeries}) then
reproduces the specified classical one-soliton solutions at $t=0$, while the
remaining terms give umbral modifications for $t\neq0$. \ 

Altogether then, we have
\begin{equation}
Q\left(  n,t=ma\right)  =q\left(  n,0\right)  +\sum_{j=1}^{m}\left(  \gamma
a\right)  ^{j}\binom{m}{j}\left(  _{\ }\Phi\left(  -\alpha e^{-\beta\left(
n+1\right)  },1-j,0\right)  -\Phi\left(  -\alpha e^{-\beta n},1-j,0\right)
_{\ }\right)  \ .
\end{equation}
These umbral results are compared to some time-continuum soliton profiles for
$t/a=0,\ 1,\ 2,\ 3,$ and $4$ in the following Figure (with $q_{0}=0$,
$\alpha=1=\beta$, and $\gamma=2\sinh\left(  1/2\right)  =1.042$ ).
%TCIMACRO{\FRAME{dtbpFU}{7.0231in}{4.6838in}{0pt}{\Qcb{Toda soliton profiles
%$q$ interpolated for all $x\in\left[  -5,5\right]  $ at integer time slices
%superimposed with their time umbral maps $Q$ (thicker curves) for $a=1$.}}%
%{}{umbraltoda.eps}{\special{ language "Scientific Word";  type "GRAPHIC";
%width 7.0231in;  height 4.6838in;  depth 0pt;  original-width 6.1678in;
%original-height 4.1053in;  cropleft "0";  croptop "1";  cropright "1";
%cropbottom "0";  filename 'UmbralToda.eps';file-properties "XNPEU";}} }%
%BeginExpansion
\begin{center}
\includegraphics[
height=4.6838in,
width=7.0231in
]%
{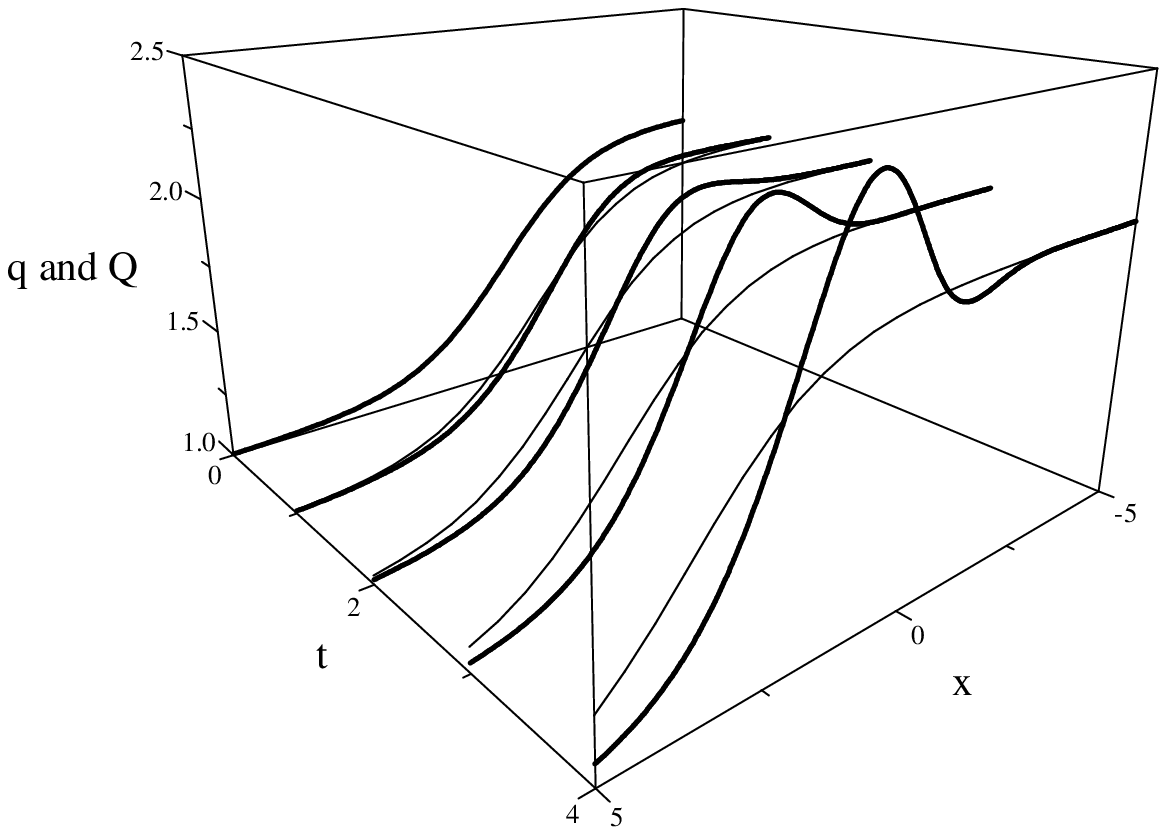}%
\\
Toda soliton profiles $q$ interpolated for all $x\in\left[  -5,5\right]  $ at
integer time slices superimposed with their time umbral maps $Q$ (thicker
curves) for $a=1$.
\end{center}
%EndExpansion

Thus, the umbral-mapped solutions no longer evolve just by translating the
profile shape. \ Rather, they develop oscillations about the classical fronts
that dramatically increase with time, that evince not only dispersion but also
generation of harmonics, and that, strictly speaking, disqualify use of the
term \href{http://www.scholarpedia.org/article/Soliton}{soliton}\ for their
description. \ Be that as it may, this model is referred to in some studies as
integrable\ \cite{Grammaticos}.

These umbral effects on wave propagation evoke scattering and diffraction by
crystals. \ But here the \textquotedblleft crystal\textquotedblright\ is
spacetime itself. \ It is tempting to speculate based on this analogy. \ In
particular, were a well-formed wave packet to pass through a localized region
of crystalline spacetime, with sufficiently large lattice spacings, the packet
could undergo dramatic deformations in shape, wavelength, and frequency ---
far greater than and very different from what would be expected just from the
dispersion of a free packet propagating through continuous space and time. \ 

\section{Concluding remarks}

We have emphasized how the umbral calculus has visibly emerged\ to provide an
elegant correspondence framework that automatically gives solutions of
ubiquitous difference equations as maps of well-known continuous functions.
\ This correspondence systematically sidesteps the use of more traditional
methods to solve these difference equations.

We have used the umbral calculus framework to provide solutions to discretized
versions of several differential equations that are widespread building-blocks
in many if not all areas of physics and engineering, thereby avoiding the
rather unwieldy frontal assaults often engaged to solve such discrete
equations directly.

We have paid special attention to the Airy, Kummer, and Whittaker equations,
and illustrated several basic principles that transform their continuum
solutions to umbral versions through the use of hypergeometric function maps.
\ The continuum limits thereof are then manifest. \ 

Finally, we have applied the solution-mapping technique to single solitons of
the Sine-Gordon, Korteweg--de Vries, and Toda systems, and we have noted how
their umbral counterparts --- particular solutions of corresponding
discretized equations --- evince dispersion and other non-solitonic behavior,
in general. Such corrections to the continuum result may end up revealing
discrete spacetime structure in astrophysical wave propagation settings.
%%THAT'S THE POINT: THERE ARE NO "THE" DISCRETIZED EQNS... THEY ARGUE ABOUT THEM.

We expect to witness several applications of the framework discussed and
illustrated here.\newpage

\section*{\protect\hypertarget{Appendix A}{Appendix A}: \ Umbral Airy
functions}

Formally, these can be obtained by expressing the Airy functions in terms of
hypergeometric functions and then umbral mapping the series. \ The continuum
problem is given by%
\begin{equation}
y^{\prime\prime}-xy=0\ \text{, }\qquad y\left(  x\right)  =C_{1}%
\operatorname{AiryAi}\left(  x\right)  +C_{2}\operatorname{AiryBi}\left(
x\right)  \ ,
\end{equation}
where%
\begin{align}
\operatorname{AiryAi}\left(  x\right)   &  =\frac{1}{3^{2/3}\Gamma\left(
2/3\right)  }\left.  _{0}F_{1}\right.  \left(  ;\frac{2}{3};\frac{1}{9}%
x^{3}\right)  -\frac{1}{3^{1/3}\Gamma\left(  1/3\right)  }\left.  _{0}%
F_{1}\right.  \left(  ;\frac{4}{3};\frac{1}{9}x^{3}\right)  \ ,\\
\operatorname{AiryBi}\left(  x\right)   &  =\frac{1}{3^{1/6}\Gamma\left(
2/3\right)  }\left.  _{0}F_{1}\right.  \left(  ;\frac{2}{3};\frac{1}{9}%
x^{3}\right)  +\frac{3^{1/6}z}{\Gamma\left(  1/3\right)  }\left.  _{0}%
F_{1}\right.  \left(  ;\frac{4}{3};\frac{1}{9}x^{3}\right)  \ .
\end{align}

The $y\longmapsto Y$ umbral images of these, solving the umbral discrete
difference equation \cite{ltw,Cholewinski03}
\begin{equation}
Y(x+2a)-2Y(x+a)+Y(x)-a^{2}xY(x+a)=0\ ,
\end{equation}
are then given by (\ref{Hyper2}) for $k=3$. \ In particular,%
\begin{align}
\operatorname*{UmAiryAi}\left(  x,a\right)   &  =\frac{1}{3^{2/3}\Gamma\left(
2/3\right)  }\left.  _{3}F_{1}\right.  \left(  -\frac{1}{3}\frac{x}{a}%
,\frac{1}{3}\left(  1-\frac{x}{a}\right)  ,\frac{1}{3}\left(  2-\frac{x}%
{a}\right)  ;\frac{2}{3};-3a^{3}\right) \nonumber\\
&  -\frac{1}{3^{1/3}\Gamma\left(  1/3\right)  }\left.  _{3}F_{1}\right.
\left(  -\frac{1}{3}\frac{x}{a},\frac{1}{3}\left(  1-\frac{x}{a}\right)
,\frac{1}{3}\left(  2-\frac{x}{a}\right)  ;\frac{4}{3};-3a^{3}\right)  \ .
\end{align}
Since the number of \textquotedblleft numerator parameters\textquotedblright%
\ in the hypergeometric function $\left.  _{3}F_{1}\right.  $ exceeds the
number of \textquotedblleft denominator parameters\textquotedblright\ by $2$,
the series expansion is at best asymptotic. \ However, the series is Borel
summable. \ In this respect, the situation is the same as for the umbral
gaussian (see Appendix B).

Alternatively, as previously mentioned in the text, using the familiar
integral representation of $\operatorname{AiryAi}\left(  x\right)  $, the
umbral map devolves to that of an exponential. \ That is to say,%
\begin{align}
\operatorname{AiryAi}\left(  xT^{-1}\right)   &  =\frac{1}{2\pi}\int_{-\infty
}^{+\infty}\exp\left(  \frac{1}{3}is^{3}+isxT^{-1}\right)  ds\\
&  \longmapsto\nonumber\\
\operatorname*{UmAiryAi}\left(  x,a\right)   &  =\frac{1}{2\pi}\int_{-\infty
}^{+\infty}e^{\frac{1}{3}is^{3}}\left(  1+isa\right)  ^{\frac{x}{a}}ds\ .
\end{align}
Just as $\operatorname{AiryAi}\left(  x\right)  $ is a real function for real
$x$, $\operatorname*{UmAiryAi}\left(  x,a\right)  $ is a real function for
real $x$ and $a$,
\begin{equation}
\operatorname*{UmAiryAi}\left(  x,a\right)  =\operatorname{Re}\left(  \frac
{1}{\pi}\int_{0}^{+\infty}e^{\frac{1}{3}is^{3}}\left(  1+isa\right)
^{\frac{x}{a}}ds\right)  \ .
\end{equation}

After some hand-crafting, the final result may be expressed in terms of just
three $\left.  _{2}F_{2}\right.  $\ generalized hypergeometric functions. \ To
wit,
\begin{align}
&  \operatorname{Re}\left(  \frac{1}{\pi}\int_{0}^{+\infty}e^{\frac{1}%
{3}is^{3}}\left(  1+isw\right)  ^{-z}ds\right) \label{2F2form}\\
&  =C_{0}\left(  w,z\right)  \times\left(  _{\ }\left.
\begin{array}
[c]{c}%
8w^{2}\sin\left(  \pi z/3\right)  ~C_{1}\left(  w,z\right)  H_{1}\left(
w,z\right)  -12w\left(  1+2\cos\left(  2\pi z/3\right)  \right)  ~C_{2}\left(
w,z\right)  H_{2}\left(  w,z\right) \\
+3~C_{3}\left(  w,z\right)  H_{3}\left(  w,z\right)
\end{array}
\right.  _{\ }\right)  \ ,\nonumber
\end{align}
where the hypergeometric functions $\left.  _{2}F_{2}\right.  \left(
a,b;c,d;z\right)  $ appear in the expression as%
\begin{align}
H_{1}\left(  w,z\right)   &  =\Gamma\left(  \frac{1}{3}z\right)  \Gamma\left(
\frac{1}{3}+\frac{1}{3}z\right)  \left.  _{2}F_{2}\right.  \left(  \frac{1}%
{3}z,\frac{1}{3}+\frac{1}{3}z;\frac{1}{3},\frac{2}{3};\frac{1}{3w^{3}}\right)
\ ,\\
H_{2}\left(  w,z\right)   &  =\Gamma\left(  \frac{1}{3}+\frac{1}{3}z\right)
\Gamma\left(  \frac{2}{3}+\frac{1}{3}z\right)  \left.  _{2}F_{2}\right.
\left(  \frac{1}{3}+\frac{1}{3}z,\frac{2}{3}+\frac{1}{3}z;\frac{2}{3},\frac
{4}{3};\frac{1}{3w^{3}}\right)  \ ,\\
H_{3}\left(  w,z\right)   &  =\Gamma\left(  \frac{2}{3}+\frac{1}{3}z\right)
\Gamma\left(  1+\frac{1}{3}z\right)  \left.  _{2}F_{2}\right.  \left(
\frac{2}{3}+\frac{1}{3}z,1+\frac{1}{3}z;\frac{4}{3},\frac{5}{3};\frac
{1}{3w^{3}}\right)  \ ,
\end{align}
and where the coefficients in (\ref{2F2form}) are%
\begin{equation}
C_{0}\left(  w,z\right)  =\frac{1}{96w^{2}}\frac{e^{-\left(  \frac{1}{3}%
\ln3+\frac{1}{2}\ln w^{2}\right)  z}}{\left(  \sin\frac{1}{3}\pi z\right)
\Gamma\left(  \frac{1}{3}z\right)  \left(  \sin\frac{1}{3}\pi\left(
1+z\right)  \right)  \Gamma\left(  \frac{1}{3}+\frac{1}{3}z\right)  \left(
\cos\frac{1}{6}\pi\left(  2z+1\right)  \right)  \Gamma\left(  \frac{2}%
{3}+\frac{1}{3}z\right)  }\ ,
\end{equation}%
\begin{align}
C_{1}\left(  w,z\right)   &  =2\sqrt[3]{3}\cos\left(  \frac{1}{6}\pi
z-\frac{1}{2}z\operatorname{signum}\left(  w\right)  \pi\right)  +\sqrt[3]%
{3}\cos\left(  \frac{1}{2}\pi z+\frac{1}{2}z\operatorname{signum}\left(
w\right)  \pi\right) \nonumber\\
&  +3^{\frac{5}{6}}\sin\left(  \frac{1}{2}\pi z+\frac{1}{2}%
z\operatorname{signum}\left(  w\right)  \pi\right)  \ ,
\end{align}%
\begin{equation}
C_{2}\left(  w,z\right)  =-\cos\left(  \frac{1}{6}\pi+\frac{1}{6}\pi
z+\frac{1}{2}\pi z\operatorname{signum}\left(  w\right)  \right)  +\sqrt
{3}\sin\left(  \frac{1}{6}\pi+\frac{1}{6}\pi z+\frac{1}{2}\pi
z\operatorname{signum}\left(  w\right)  \right)  \ ,
\end{equation}%
\begin{align}
C_{3}\left(  w,z\right)   &  =3\sqrt[6]{3}\cos\left(  \frac{1}{2}\pi
z-\frac{1}{2}\pi z\operatorname{signum}\left(  w\right)  \right)
-6\sqrt[6]{3}\cos\left(  \frac{1}{6}\pi z+\frac{1}{2}\pi
z\operatorname{signum}\left(  w\right)  \right) \nonumber\\
&  +6\sqrt[6]{3}\cos\left(  \frac{5}{6}\pi z+\frac{1}{2}\pi
z\operatorname{signum}\left(  w\right)  \right)  +3\times3^{\frac{2}{3}}%
\sin\left(  \frac{1}{2}\pi z-\frac{1}{2}\pi z\operatorname{signum}\left(
w\right)  \right) \nonumber\\
&  -3\sqrt[6]{3}\cos\left(  \frac{1}{2}\pi z-\frac{1}{2}\pi
z\operatorname{signum}\left(  w\right)  \right)  +2\times3^{\frac{2}{3}}%
\sin\left(  \frac{1}{6}\pi z+\frac{1}{2}z\pi\operatorname{signum}\left(
w\right)  \right) \nonumber\\
&  +3^{\frac{2}{3}}\sin\left(  \frac{1}{2}\pi z+\frac{1}{2}\pi
z\operatorname{signum}\left(  w\right)  \right)  +2\times3^{\frac{2}{3}}%
\sin\left(  \frac{5}{6}\pi z+\frac{1}{2}\pi z\operatorname{signum}\left(
w\right)  \right)  \ .
\end{align}
While the coefficient functions $C_{0-3}$ are not pretty, they are comprised
of elementary functions, and they are nonsingular functions of $z$. \ \ On the
other hand, the hypergeometric functions do have singularities and
discontinuities for negative $z$. \ However, the net result for
$\operatorname*{UmAiryAi}$ is reasonably well-behaved.

We plot $\operatorname*{UmAiryAi}\left(  x,a\right)  $ for $a=0,\ \pm\frac
{1}{4},\ \pm\frac{1}{2},$ and $\pm1$.%
%TCIMACRO{\FRAME{dtbpFU}{5.7804in}{3.8545in}{0pt}%
%{\Qcb{$\operatorname*{UmAiryAi}\left(  x,a\right)  $ for $a=\pm1$, $\pm1/2$,
%and $\pm1/4$ (red, blue, \& green dashed/solid curves, resp.) compared to
%$\operatorname{AiryAi}\left(  x\right)  =\operatorname*{UmAiryAi}\left(
%x,0\right)  $\ (black curve).}}{}{umbralairy__4.eps}%
%width 5.7804in;  height 3.8545in;  depth 0pt;  original-width 7.6828in;
%original-height 5.1125in;  cropleft "0";  croptop "1";  cropright "1";
%cropbottom "0";  filename 'UmbralAiry__4.eps';file-properties "XNPEU";}} }%
%BeginExpansion
\begin{center}
\includegraphics[
height=3.8545in,
width=5.7804in
]%
{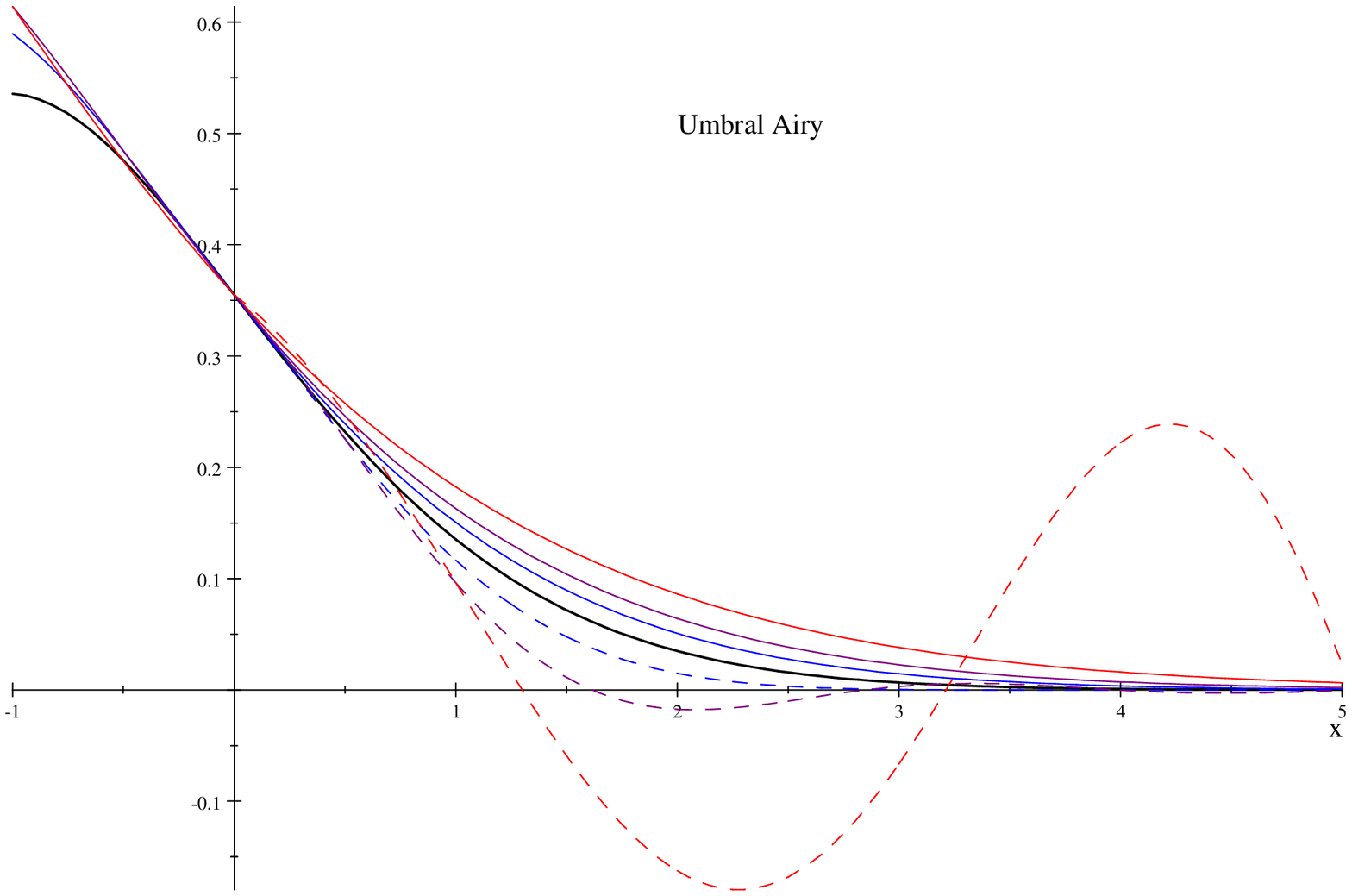}%
\\
$\operatorname*{UmAiryAi}\left(  x,a\right)  $ for $a=\pm1$, $\pm1/2$, and
$\pm1/4$ (red, blue, \& green dashed/solid curves, resp.) compared to
$\operatorname{AiryAi}\left(  x\right)  =\operatorname*{UmAiryAi}\left(
x,0\right)  $\ (black curve).
\end{center}
%EndExpansion
%TCIMACRO{\FRAME{dtbpFU}{5.7804in}{3.8536in}{0pt}{\Qcb{{}}}{}%
%{umbralairy__5.eps}{\special{ language "Scientific Word";  type "GRAPHIC";
%width 5.7804in;  height 3.8536in;  depth 0pt;  original-width 7.6828in;
%original-height 5.1125in;  cropleft "0";  croptop "1";  cropright "1";
%cropbottom "0";  filename 'UmbralAiry__5.eps';file-properties "XNPEU";}} }%
%BeginExpansion
\begin{center}
\includegraphics[
height=3.8536in,
width=5.7804in
]%
{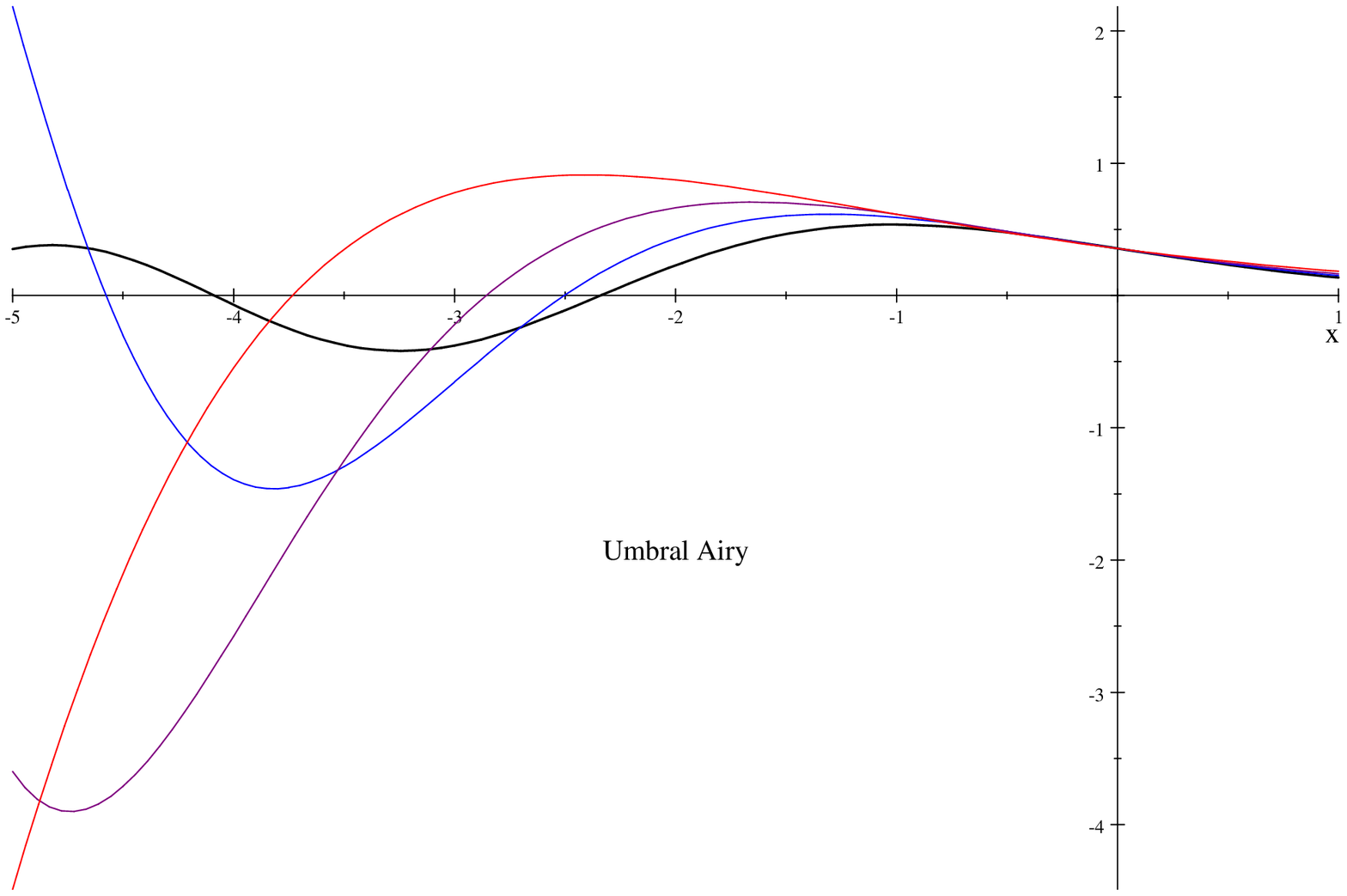}%
\\
{}%
\end{center}
%EndExpansion

\section*{\protect\hypertarget{Appendix B}{Appendix B}: \ Umbral gaussians}

As discussed in the text, straightforward discretization of the series yields
the umbral gaussian map:%
\begin{gather}
e^{-x^{2}}\longmapsto G\left(  x,a\right)  =\sum_{n=0}^{\infty}\frac
{(-)^{n}[x]^{2n}}{n!}=\sum_{n=0}^{\infty}\frac{(-)^{n}}{n!}~x(x-a)\cdots
(x-\left(  2n-1\right)  a)\label{UmbralGauss}\\
=\left.  _{2}F_{0}\right.  \left(  -\frac{1}{2}\frac{x}{a},\frac{1}{2}\left(
1-\frac{x}{a}\right)  ;-4a^{2}\right)  \ . \label{2F0UmbralGauss}%
\end{gather}
(NB \ $G\left(  x,a\right)  \neq G\left(  -x,a\right)  $.) \ Now, it is clear
that term by term the series (\ref{UmbralGauss}) reduces back to the continuum
gaussian as $a\rightarrow0$. \ Nonetheless, since the series is asymptotic and
not convergent for $\left\vert a\right\vert >0$, it is interesting to see how
this limit is obtained from other representations of the hypergeometric
function in (\ref{2F0UmbralGauss}), in particular from using readily available
numerical routines to evaluate $\left.  _{2}F_{0}\right.  $ for specific small
values of $a$. \ Some examples are shown here. \
%2F0 computed using MuPAD numerical routines%
%TCIMACRO{\FRAME{dtbpFU}{6.6227in}{4.4105in}{0pt}{\Qcb{$G\left(  x,1/2^{n}%
%\right)  $ versus $x\in\left[  -3,2\right]  $, for $n=1,\ 2,\ $and $3$, in
%red, blue, and green, respectively, compared to $G\left(  x,0\right)
%=\exp\left(  -x^{2}\right)  $, in black.}}{}{umbralgaussian.eps}%
%width 6.6227in;  height 4.4105in;  depth 0pt;  original-width 6.2759in;
%original-height 4.1701in;  cropleft "0";  croptop "1";  cropright "1";
%cropbottom "0";  filename 'UmbralGaussian.eps';file-properties "XNPEU";}} }%
%BeginExpansion
\begin{center}
\includegraphics[
height=4.4105in,
width=6.6227in
]%
{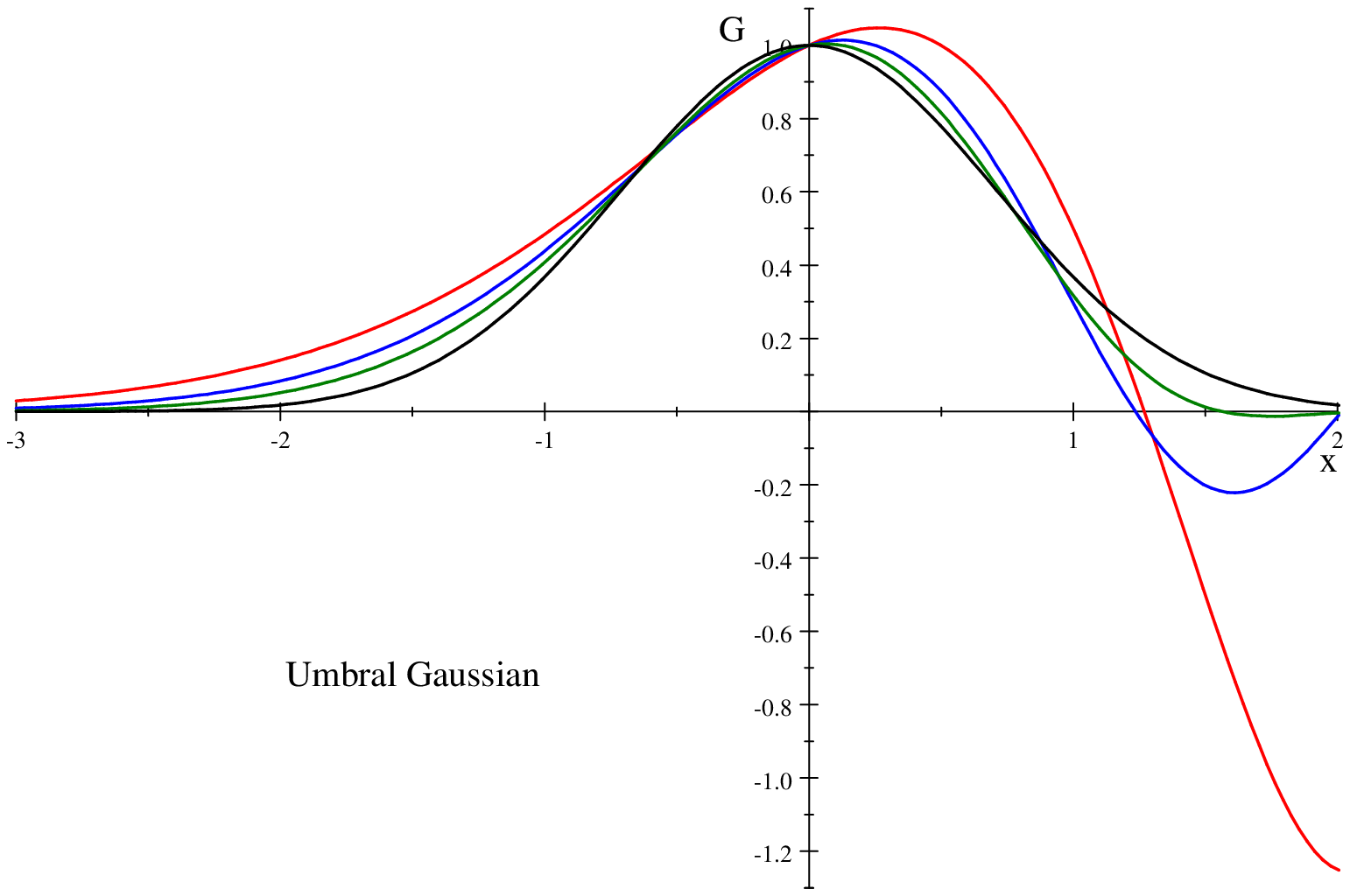}%
\\
$G\left(  x,1/2^{n}\right)  $ versus $x\in\left[  -3,2\right]  $, for
$n=1,\ 2,\ $and $3$, in red, blue, and green, respectively, compared to
$G\left(  x,0\right)  =\exp\left(  -x^{2}\right)  $, in black.
\end{center}
%EndExpansion

Mathematica$^{\textregistered}$ code is available
\href{http://server.physics.miami.edu/~curtright/UmbralGaussianMath.html}{online}
to produce similar graphs, for those interested. \ It is amusing that
Mathematica manipulates the Borel regularized sum to render the $\left.
_{2}F_{0}\right.  $ in question in terms of Tricomi's confluent hypergeometric
function $U$, as discussed above in the context of Kummer's equation, cf.
(\ref{HypergeometricU}). \ Thus $G$ can also be expressed in terms of $\left.
_{1}F_{1}\right.  $s. \ The relevant identities are:%
\begin{equation}
G\left(  x,a\right)  =\left(  2a\right)  ^{\frac{x}{a}-1}U\left(  \frac{1}%
{2}\left(  1-\frac{x}{a}\right)  ,\frac{3}{2},\frac{1}{4a^{2}}\right)
=\sqrt{\pi}\left(  2a\right)  ^{\frac{x}{a}}\left(  \frac{\left.  _{1}%
F_{1}\right.  \left(  -\frac{1}{2}\frac{x}{a};\frac{1}{2};\frac{1}{4a^{2}%
}\right)  }{\Gamma\left(  \frac{1}{2}\left(  1-\frac{x}{a}\right)  \right)
}-\frac{\left.  _{1}F_{1}\right.  \left(  \frac{1}{2}\left(  1-\frac{x}%
{a}\right)  ;\frac{3}{2};\frac{1}{4a^{2}}\right)  }{a~\Gamma\left(  -\frac
{1}{2}\frac{x}{a}\right)  }\right)  \ .
\end{equation}
\newpage

\noindent\textbf{Acknowledgements:}

\noindent\textsl{{\small This work was supported in part by NSF Award
PHY-1214521; and in part, the submitted manuscript has been created by
UChicago Argonne, LLC, Operator of Argonne National Laboratory. Argonne, a
U.S. Department of Energy Office of Science laboratory, is operated under
Contract No. DE-AC02-06CH11357. The U.S. Government retains for itself, and
others acting on its behalf, a paid-up nonexclusive, irrevocable worldwide
license in said article to reproduce, prepare derivative works, distribute
copies to the public, and perform publicly and display publicly, by or on
behalf of the Government. \ TLC was also supported in part by a University of
Miami Cooper Fellowship. \ }}

\end{document}